\documentstyle[12pt]{article}
\normalsize
\let\oldtheequation=\theequation
\def\doteqs#1{\setcounter{equation}{0}
            \def\theequation{{#1}.\oldtheequation}}
\newcounter{sxn}
\def\sx#1{\addtocounter{sxn}{1} \bigskip\medskip \goodbreak
\noindent{\large\bf\centerline{\thesxn.~~#1}} \nobreak \medskip}
\def\sxn#1{\sx{#1} \doteqs{\thesxn}}

\newcounter{axn}

\date{}

\begin{document}
\bibliographystyle{unsrt}
\footskip 1.0cm
\thispagestyle{empty}
\setcounter{page}{0}
\begin{flushright}
PSU/TH/144\\
SU-4240-578\\
November, 1994 \\
\end{flushright}
\vspace{10mm}

\centerline {\LARGE Duality and the Fractional Quantum Hall Effect}
\vspace*{15mm}
\centerline {\large  A.P.Balachandran$^{1}$, L. Chandar$^{1}$,
B.Sathiapalan$^{2}$}

\vspace*{5mm}
\centerline {\it $^{1}$ Department of Physics, Syracuse University,}
\centerline {\it Syracuse, NY 13244-1130, U.S.A.}
\centerline {\it $^{2}$ Department of Physics, Pennsylvania State University,}
\centerline {\it 120, Ridge View Drive, Dunmore, PA-18512, U.S.A.}
\vspace*{15mm}
\normalsize
\centerline {\bf Abstract}
\vspace*{5mm}

The edge states of a sample displaying the quantum Hall effect (QHE)
can be  described by a 1+1 dimensional (conformal) field theory of $d$ massless
scalar fields taking values on a $d$-dimensional torus.  It is known from the
work of Naculich,
Frohlich et al.\@ and others that the  requirement
of chirality of currents in this \underline{scalar} field theory
implies the Schwinger anomaly in the presence of an electric field,
the anomaly coefficient being related in a specific way to Hall
conducvivity.  The latter can take only certain restricted values with
odd denominators if the theory admits fermionic states.
We show that the duality symmetry under the $O(d,d;{\bf Z})$ group of the free
theory
transforms the Hall conductivity in a well-defined way and relates
integer and fractional QHE's.  This means, in particular, that the edge
spectra for dually related Hall conductivities are identical, a prediction
which may be experimentally testable.
We also show that Haldane's hierarchy as well as certain of Jain's fractions
can be
reproduced from the Laughlin fractions using the duality transformations.  We
thus find a framework for a unified description of the QHE's occurring at
different fractions.
We also give a simple derivation of the wave functions for fractions in
Haldane's hierarchy.

\newpage

\baselineskip=24pt
\setcounter{page}{1}
\newcommand{\be}{\begin{equation}}
\newcommand{\ee}{\end{equation}}
\newcommand{\Gij }{\mbox{$G^{ij}$}}
\newcommand{\Bij }{\mbox{$B^{ij}$}}
\newcommand{\Am }{\mbox{$A _{\mu}$}}
\newcommand{\Ai }{\mbox{$A ^{i}$}}
\newcommand{\Aj }{\mbox{$A ^{j}$}}
\newcommand{\alm }{\mbox{$\alpha _{\mu}$}}
\newcommand{\ali }{\mbox{$\alpha _{i}$}}
\newcommand{\alj }{\mbox{$\alpha _{j}$}}
\newcommand{\am }{\mbox{$a _{\mu}$}}
\newcommand{\An }{\mbox{$A _{\nu}$}}
\newcommand{\aln }{\mbox{$\alpha _{\nu}$}}
\newcommand{\an }{\mbox{$a _{\nu}$}}
\newcommand{\Al }{\mbox{$A _{\lambda}$}}
\newcommand{\all }{\mbox{$\alpha _{\lambda}$}}
\newcommand{\al }{\mbox{$a _{\lambda}$}}
\newcommand{\dm }{\mbox{$\partial _{\mu}$}}
\newcommand{\dn }{\mbox{$\partial _{\nu}$}}
\newcommand{\dl }{\mbox{$\partial _{\lambda}$}}
\newcommand{\pa }{\mbox{$\partial $}}
\newcommand{\ffi }{\mbox{$\phi ^{i} $}}
\newcommand{\ffj }{\mbox{$\phi ^{j} $}}
\newcommand{\qi }{\mbox{$ q ^{i} $}}
\newcommand{\qj }{\mbox{$ q ^{j} $}}
\newcommand{\emnl }{\mbox{$\epsilon _{\mu \nu \lambda}$}}
\newcommand{\efp }{\mbox{$\frac{e^{2}}{4 \pi}$}}
\newcommand{\esp }{\mbox{$\frac{e^{2}}{2 \pi}$}}

\newcommand{\e  }{\mbox{$\epsilon $}}

\sxn{Introduction}

Fractional Quantum Hall Effect (FQHE) \cite{hall} is the phenomenon of
quantized conductance at values that are
fractions of what simple considerations suggest for a system of non-interacting
electrons.
There have been many attempts to understand this phenomenon.  Following the
original work of Laughlin \cite{laugh} describing
the ground state of the FQHE for a filling fraction $\nu =\frac{1}{m}\;$ (where
$m$ is odd), there have been extensions describing the FQHE for a general
rational $\nu$ \cite{hald,halp}.  The theory of Jain \cite{jain} has the
extra appeal
of establishing a connection between the integer and fractional effects.  This
is important phenomenologically (apart from being aesthetically appealing),
since experimentally there does not seem to be much difference between the
physics
at Hall plateaus corresponding to integer and fractional filling factors.

Many of these models can be described by Chern-Simons theories
\cite{sri,read,zee,froh} in
the interior of the disc.  The advantage of using Chern-Simons theories is
that they have observables only at the boundary of the manifold \cite{witt,bal}
(here a
disc).  In particular, these theories assert that quantum Hall effect can be
described by observables living at the edge of the sample.  A simple way to
motivate these edge currents is from classical considerations.  An edge for the
Hall sample arises because of the existence of a potential barrier that
prevents the electrons from escaping the finite region.  We thus expect a
radially outward electric field at the boundary.  This will now create a Hall
current tangential to the boundary because of the presence of a magnetic field
perpendicular to the sample.  This Hall current is the edge current mentioned
above.
This argument can be made precise by considering the quantum mechanical wave
functions of the electrons confined by a potential barrier as has been done by
Halperin \cite{halpe}.  In this work he has also shown the intimate connection
between this current and the integer quantization of Hall conductivity.
There have subsequently been many papers emphasizing the importance of
edge states in understanding aspects
of both Integer Quantum Hall Effect (IQHE) and FQHE
\cite{wen,wil,froh,dim,kar,cap}.

In this work, we utilize the fact that the Chern-Simons theories mentioned
above are equivalent to theories of chiral bosons living at the edge of the
sample
\cite{witt,bal}.  We thus use a 1+1 dimensional theory consisting of
$d$ massless scalar fields with the target manifold being a $d$-torus.  We can
then write down an action similar to that used in various dimensional
compactification schemes for string theories \cite{nar,gre}.  These theories
have
certain duality symmetries \cite{kik,bala,dua} which in the
context of Hall effect, have interesting
consequences.  When we couple the system to electromagnetism, it is seen that
the requirement of chirality of the fields and currents relates the Hall
conductivity to the so-called
``metric'' on the target torus \cite{zee,froh,chan}.  If we assume that the
scalar
fields have definite electromagnetic couplings, then, for different choices of
the metric, we get different Hall conductivities. Now, with the help
of the duality symmetry, we can relate integer and fractional values, and by
choosing specific elements of the generalised
duality transformations, we can get the hierarchical values predicted by
Haldane as well as Jain's fractions.  In this way, we can relate the integer
and fractional conductivities in these schemes.  {\em The important
prediction that arises as a consequence is that the spectra of edge
excitations are identical for the two conductivities so related.}  This
prediction may be experimentally testable.

Using
conformal field theory techniques, we also give a simple
derivation of the
hierarchical wave functions for electrons and quasiparticles that have appeared
in the literature \cite{hald,halp,fv,fub,nrea,cris}.  We will see later in
Sections 3
and 4 that this prescription works only so long as we are in the lowest Landau
level.

An important result we have used in the course of this work is due to
Naculich \cite{nac} (and also to Frohlich et al. \cite{froh}, Wilczek
\cite{wil}
and others).  It is
that the imposition of chirality on the above \underline{scalar} field
theory leads to the Schwinger anomaly \cite{sch} (the anomaly
coefficient being related in a known way \cite{call,nac} to Hall
conductivity).

In Section 2, we specialize to the case of one scalar field with its
radius $R$ of compactification held arbitrary to begin with.
By imposing chirality and requiring the existence of fermionic states in the
theory, we get restrictions on
this radius.  We then gauge this theory preserving chirality, and show
how this radius of
compactification is related to Hall conductivity.  By using
the standard $R\rightarrow \frac{1}{R}$ duality that exists for this theory,
we
then relate Hall conductivities at two fractions which are inverses of each
other.  By allowing many scalar fields at the boundary, we then show that the
fractions compatible with the restrictions obtained on $R$ are
always those with odd denominators.  These fields incidentally can be
interpreted in terms of fields for the electron and the flux tubes attached to
it.

In Section 3, we study the response of Hall conductivity to the
more general $O(d,d;{\bf Z})$ ``symmetry'' that exists for the free field
theory
with many
scalar fields.  We will see that this  symmetry makes the
problem much richer but also much more complicated.

In Section 4, we motivate a choice of the internal
``metric'' for the scalar field theory.  Then
we show how we can obtain both Haldane's continued fractions as well as
certain of Jain's fractions from subsets of the $O(d,d;{\bf Z})$
transformations of Section 3.

In Section 5, we present some conclusions and describe a few open questions.

\sxn{Duality and FQHE for a Chiral Boson}

{\bf 2.1. The Chiral Constraint on the Left-Right Symmetric Boson}
\nopagebreak
\par
We will imagine for the rest of the paper that the Hall system is on a
disc $D$ with a circular boundary $\partial D$.  In this situation, the
excitations at the edge are described by massless fields in 1+1 dimensions.
They give the edge currents mentioned in Section 1.  Since a
fermionic theory in 1+1
dimensions can always be bosonised, we may as well work with scalar
fields.  To begin with, we
assume that the theory is a free theory with just one scalar field.  We assume
that it is valued in a circle.
The action for this theory is
\be
A_{0}= \frac{R^{2}}{8\pi}\int dt\int _{0}^{2\pi}dx (\partial _{\mu}\phi )
(\partial ^{\mu}\phi )\label{2.1}
\ee
where our spacetime metric has diagonal elements ($1,-1$) [and zero elsewhere]
and $\phi (x)$ is identified with $\phi (x) + 2\pi$.

In the above,
we have set the radius of the disc (not to be confused with the radius of
compactification in the target space of the scalar field) equal to 1.
Restoring the actual
radius will rescale the spectrum of the Hamiltonian of the above action by a
factor $1/\mbox{(radius) }$.

Note that by assumption,
$e^{i\phi}$ is valued in a circle, $e^{i\phi (x)}\in S^{1}$.  The latter
condition enables the existence of
solitons, similar to sine-Gordon solitons, in this theory.  As usual
some of these solitons will later on be
interpreted
as  fermions.

The coefficient $R^{2}$ outside the integral in (\ref{2.1}) is, as
of now, arbitrary.

The above action by itself defines a left-right symmetric theory whereas
the real Hall system describes a situation which is chiral.  We thus need to
impose chirality as a constraint on the states of the theory described by
(\ref{2.1}).  If we choose to impose the constraint that the field is
left-moving, then
\be
\partial _{-}\phi \equiv \frac{1}{2}(\dot{\phi}-\phi ')=0.\label{2.2}
\ee
[Here and below, we define $x_{\pm}=t\pm x$.  So $dx^{\mu}\partial _{\mu}
=dx^{+}\partial _{+} +dx^{-}\partial _{-}$ where $\partial _{\pm}=\frac{1}{2}
(\partial _{0}\pm \partial _{1})$, and $A_{\mu}dx^{\mu}=A_{+}dx^{+}+A_{-}dx^{-}
$ where $A_{\pm}=\frac{1}{2}(A_{0}\pm A_{1})$.  Also $\partial _{\pm}$ refer to
differentiations with $x_{\pm}$ as independent variables while $\partial _{0},
\partial _{1}$ refer to differentiations with $t$ and $x$ as independent
variables.]  We should then check if such a constraint is compatible
with the equation of
motion that arises from the action (\ref{2.1}).  This equation of motion is
\be
\partial _{\mu}\partial ^{\mu} \phi \equiv \frac{1}{4}\partial _{+}\partial
_{-}\phi
=0,\label{2.25}
\ee
which is clearly compatible with (\ref{2.2}).

But this will no longer be true
when we gauge the action (\ref{2.1}) with an external electromagnetic field
described by the potential $A_{\mu}$ to obtain
\begin{eqnarray}
S_{0}[\phi ,A] &=& \frac{R^{2}}{8\pi}\int d^{2}x(D_{\mu}\phi )
^{2}-\frac{1}{4k^{2}}\int d^{2}xF_{\mu\nu}F^{\mu\nu}, \nonumber\\
D_{\mu}\phi &\equiv &\partial _{\mu}\phi -eA_{\mu}, \label{2.23}
\end{eqnarray}
$k$ being a constant and $F_{\mu\nu}=\partial _{\mu}A_{\nu}-
\partial _{\nu}A_{\mu}$.  [The field $A_{\mu}$ is of course an addition to the
vector potential describing the
magnetic field ever present on the Hall sample and responsible for instance
for the Landau levels.]
This gauging has been done in accordance with the gauge transformation law
\begin{eqnarray}
e^{i\phi} & \rightarrow & e^{i(\phi +e\epsilon )},\nonumber\\
A_{\mu} & \rightarrow & A_{\mu}+\partial _{\mu}\epsilon \label{2.24}
\end{eqnarray}
so that $e^{i\phi}$ transforms like the phase of a charged Higgs field.

The equations of motion of the above action are
\begin{eqnarray}
&&\frac{1}{k^{2}}\partial ^{\nu}F_{\nu\mu} = \frac{eR^{2}}{4\pi}(D_{\mu}\phi
),\nonumber\\
&&\partial ^{\mu}(D_{\mu}\phi )=  0   .
\label{2.26}
\end{eqnarray}

The constraint (\ref{2.2}) for the ungauged theory should now be replaced by
the gauge invariant constraint
\be
D_{-}\phi \equiv (\partial _{-}\phi -eA
_{-}) = 0.\label{2.27}
\ee
However this constraint turns out to be incompatible with
equation (\ref{2.26}) above.  This can be seen as follows:
\begin{eqnarray}
0=\partial _{+}D_{-}\phi & = & \frac{1}{4}(\partial _{0} +\partial
_{1})(D_{0}\phi
-D_{1}\phi )\nonumber\\
\Rightarrow \partial _{\mu}(\frac{eR^{2}}{4\pi}D^{\mu}\phi) & = &
-\frac{e^{2}R^{2}}{4\pi}E, \nonumber\\
E&:=&\partial _{0}A_{1}-\partial _{1}A_{0}.\label{2.29}
\end{eqnarray}
Thus the gauge invariant chirality constraint leads to an equation inconsistent
with the last equation in (\ref{2.26}).  Note that equation (\ref{2.29}) is
identical to the Schwinger anomaly equation for the $U(1)$ current of
a chiral fermion \cite{sch}.
But we have obtained it here without appealing to the existence of fermions.
The {\it only} input that we need is {\it chirality}. We thus arrive at the
interesting result that
chirality in a bosonic theory can lead to
anomalies.   Moreover, the anomaly coefficient is determined completely by the
constant $R^{2}$. [In our context, these results are originally due to Naculich
\cite{nac}.  See also Frohlich \cite{froh}, Wilczek \cite{wil} and
Alvarez-Gaume and Witten \cite{lag}.]

Let us return to (\ref{2.29}).  To obtain this equation,
the action (\ref{2.23}) has to be augmented to the new action
\begin{eqnarray}
&&S(\phi , A)= S_{0}(\phi , A) +\frac{eR^{2}}{4\pi}\int d\phi A \equiv
S_{0}(\phi ,A)+\frac{eR^{2}}{4\pi}\int d^{2}x \epsilon ^{\mu\nu}\partial
_{\mu}\phi A_{\nu},\nonumber\\
&&\epsilon ^{\mu\nu}=-\epsilon ^{\nu\mu},\;\;\; \epsilon ^{01}=1, \label{anact}
\end{eqnarray}
 so that its equations of motion are now
compatible with the chirality constraint.  The current $j^{\mu}$ for
this theory is
\be
j^{\mu} \equiv -\frac{\delta }{\delta A_{\mu}}[S(\phi ,A)+\frac{1}{4k^{2}}\int
d^{2}x F_{\mu\nu}F^{\mu\nu}]= \frac{eR^{2}}{4\pi}
(D^{\mu}\phi +\epsilon ^{\mu\nu}\partial _{\nu}\phi ) \label{2.295}
\ee
and its equation of motion gives the anomaly equation
\be
\partial _{\mu}j^{\mu} =-\frac{e^{2}R^{2}}{4\pi}E. \label{naivan}
\ee

The current $j^{\mu}$ is not gauge invariant.
But we want to identify the current at the edge with the electromagnetic
current there in a Hall sample.  It should therefore be gauge invariant unlike
$j^{\mu}$.  It should also be chiral whereas $j_{-}=j_{0}-j_{1}$ is non-zero.
  Thus $j^{\mu}$ cannot be the electromagnetic current at the
edge.

We can however overcome both these drawbacks of $j^{\mu}$ by replacing
$\partial _{\nu}$ in (\ref{2.295}) by the covariant derivative $D_{\nu}$.  We
then get the current
\be
{\cal J}^{\mu} = \frac{eR^{2}}{4\pi}(D^{\mu}\phi +\epsilon
^{\mu\nu}D_{\nu}\phi ).\label{2.296}
\ee
This is the current with the ``covariant'', and not the ``consistent'' anomaly
\cite{zum,nac}.

There is a very nice interpretation for such a
modification of the current
from $j^{\mu}$ to ${\cal J}^{\mu}$ which also shows why ${\cal J}^{\mu}$ is the
physical edge current. (Similar arguments have appeared in \cite{froh}.)  This
is as follows.   The modified action $S(\phi ,A)$
which contains also the anomaly term is clearly not invariant under the
gauge transformations (\ref{2.24}).  The physical reason for this is
the fact that the
boundary of the Hall sample is not isolated.  Thus there can be a flow of
charge from (to) the edge to (from) the interior.  We therefore do not expect
the charge at the boundary to be conserved by itself.  But since we do know
that the total charge (which includes the charge in the interior of
the disc as well) is conserved, the action describing dynamics in the bulk
should change
under the above gauge transformations (\ref{2.24}) in such a way that the total
action is gauge invariant \cite{call,nac,froh,wil}.   The simplest such action
in the bulk is the Chern-Simons term
\begin{eqnarray}
&&-\frac{e^{2}R^{2}}{4\pi}\int _{D\times {\bf R}^{1}}AdA \equiv -\frac{e^{2}
R^{2}}{4\pi}\int _{D\times {\bf R}^{1}}
d^{3}x\epsilon ^{\mu\nu\lambda}A_{\mu}\partial _{\nu}A_{\lambda}
,\nonumber\\
&& A:=A_{\mu}dx^{\mu},\;\; \epsilon ^{\mu\nu\lambda}=\mbox{ Levi-Civita symbol
with }\epsilon ^{0r\theta }=1, \label{new 2}
\end{eqnarray}
(apart from other gauge invariant terms
like the Maxwell term). [Here {\bf R}$^{1}$ accounts for time.  Also wedge
symbols between differential forms will be omitted in this paper.]  Since
\be
\label{csv}
\delta \int _{D\times {\bf R}^{1}} AdA =2\int _{D\times {\bf R}^{1}}dA\; \delta
A -\int _{\partial D\times {\bf R}^{1}}A\delta A,
\ee
we see that the current that we
obtain from this action has a bulk piece as well as an edge piece.  Also the
latter
contribution is precisely the one that changes $j^{\mu}$ to ${\cal J}^{\mu}$
\cite{nac}!

It is interesting to note that the full action (that is, the action in the
interior plus the action at the boundary) can be written in the
manifestly gauge invariant form \cite{wil}
\begin{eqnarray}
S_{total}&=& S_{0} -\frac{R^{2}}{4\pi}\int _{D\times {\bf R}^{1}}D\phi d(D\phi
),\nonumber\\
D\phi d(D\phi )&\equiv & d^{3}x\epsilon ^{\mu\nu\lambda}(D_{\mu}\phi
)\partial _{\nu}(D_{\lambda}\phi ). \label{2.2965}
\end{eqnarray}

Now, using the equations of motion of the action (\ref{anact}), we
get the anomaly equation
\be
\partial _{\mu}{\cal J}^{\mu} = -\frac{e^{2}R^{2}}{2\pi}E \label{2.297}
\ee
Note that the anomaly term in (\ref{2.297}) is \underline{twice} that in
(\ref{naivan}).

On the other hand, using arguments as in \cite{call,fend} (see also
\cite{chan}), we know
that the anomaly coefficient is the same as the Hall conductivity
$\sigma _{H}$.

Let us recapitulate this argument briefly \cite{call,chan}.  The anomaly
equation (\ref{2.297}) tells us that the charge is not conserved.  In fact the
rate of change of charge at the boundary is $-\frac{e^{2}R^{2}}{2\pi}\int dx\;
E$.  Now, a Hall system obeys the Hall equation in the bulk which
in particular gives rise to the equation $J^{r}=-\sigma _{H}E_{\theta}$ in the
bulk (where
$r,\theta$ refer to polar coordinates on the disk).  From this equation we get
the rate of change of charge in the bulk to be equal to $\sigma _{H}\int
_{\partial D}dx\; E
$.  Since charge that is escaping from bulk has to go to the boundary,
(\ref{2.297}) gives the relation
\be
\sigma _{H} =\frac{e^{2}R^{2}}{2\pi}. \label{2.298}
\ee

It is satisfying to note that this is the same as the Hall conductivity
that one would have obtained from the Chern-Simons action (\ref{new 2}) (along
with
the Maxwell term) in the interior. Thus, for the action (\ref{new 2}), the
current in the bulk is
\be
J^{\mu}=\frac{e^{2}R^{2}}{2\pi}\epsilon ^{\mu\nu\lambda}\partial
_{\nu}A_{\lambda}, \label{bulkJ}
\ee
from which we get
\begin{eqnarray}
&&J^{i}=-\frac{e^{2}R^{2}}{2\pi}\epsilon ^{ij}E_{j}, \nonumber \\
&&\epsilon ^{r\theta}=-\epsilon ^{\theta r}=1,\label{Halleq}
\end{eqnarray}
which is exactly the statement that the Hall conductivity is
$\frac{e^{2}R^{2}}{2\pi}$.

Let us now return to the gauge invariant constraint (\ref{2.27}).  We will now
see that elementary quantum theory imposes non-trivial restrictions on the
allowed $R^{2}$ because of this constraint.  Firstly, we have as a consequence
of (\ref{2.27}),
\be
\int _{0}^{2\pi} dx (D_{0}\phi -D_{1}\phi ) = 0.\label{2.3}
\ee

Let
\be
\Pi =\frac{R^{2}}{4\pi}(D_{0}\phi +eA_{1}) \label{new 3}
\ee
 denote the momentum density conjugate to
$\phi$.  We then have the equal time commutation relations (CR's)
\begin{eqnarray}
[\phi (x),\phi (x')]&=&0,\nonumber\\
{[}\Pi (x),\Pi (x'){]}&=&0,\nonumber\\
{[}\phi (x),\Pi (x'){]}&=&i\delta (x-x'). \label{2.31}
\end{eqnarray}
[Often we will not indicate the time-dependence of the fields and their modes
as they
will all be at equal times.]
In terms of $\Pi$, (\ref{2.3}) reads
\begin{eqnarray}
&&\int _{0}^{2\pi}dx (\frac{4\pi\Pi }{R^{2}}-\phi ') =0,\nonumber\\
&&\phi ':=\partial _{1}\phi .\label{2.4}
\end{eqnarray}

We next briefly analyse the properties of $\Pi$ and $\phi '$.

{}From the CR of $\phi$ and $\Pi$, we get
\be
[\phi (x),\int dx' \Pi (x')] = i \label{2.6}
\ee
at equal times.
Thus $\int dx'\Pi (x')$ is canonically conjugate to $\phi$, or rather
to the
spatially constant mode $\phi _{0}$ of $\phi$.

Now since $\phi (x)$
is identified with $\phi (x) +2\pi$, the dependence of
wave functions on $\phi
_{0}$ should satisfy
\begin{eqnarray}
\psi (\phi _{0} +2\pi ) & = & e^{i2\pi\alpha}\psi (\phi _{0}),\nonumber\\
\psi '(\phi _{0} +2\pi ) & = & e^{i2\pi\alpha}\psi '(\phi _{0})\label{2.7}
\end{eqnarray}
where $\alpha$ is a real number taking values in [0,1[.
(Only the dependence of $\psi$ on $\phi _{0}$ has been displayed
here.)  As is well
known, $\alpha$ can be interpreted in terms of a flux passing through the
circle in the
target space (the circle on which the field $\phi$ takes values).

It follows
from (\ref{2.7}) that
\be
\psi _{m}(\phi _{0}) = e^{i(m+\alpha )\phi _{0}},\;\;\; m=0,\pm 1,\pm
2,\ldots \label{2.8}
\ee
are the eigenfunctions of
\be
p:=\int dx\Pi (x)\equiv -i\frac{\partial}{\partial \phi _{0}} \label{new 4}
\ee
while the
corresponding eigenvalues are $m+\alpha$.  Therefore
\be
\mbox{Spec }p \equiv \mbox{ Spectrum of }p =\{ m+\alpha \; :m\in {\bf Z}
\} .\label{2.9}
\ee

Also, the identification of $\phi (x)$ with $\phi (x)+2\pi$ gives rise to the
following condition on $\int dx\phi '(x)$:
\be
\int dx\phi '(x) = 2\pi N,\;\;\; N\in {\bf Z}.\label{2.10}
\ee

In our current approach, there is no operator in the theory which can
change the winding number.  We can therefore regard all states as
having a fixed winding number $N$, different choices of $N$ giving
different quantisations of (\ref{2.1}).

With $N$ fixed, it follows
from (\ref{2.4}) that $p$ too has a fixed value $(m+\alpha )$ in a
given quantisation where
\[ \frac{4\pi (m+\alpha )}{R^{2}} -2\pi N  = 0.\]
It gives, for $N\neq 0$,
\be
R^{2}  = \frac{2(m+\alpha )}{N}.\label{2.11}
\ee
Since $R^{2}>0$, this implies in particular that
\be
\frac{m+\alpha }{N}>0 \mbox{ if } N\neq 0, \label{new 5}
\ee
while if $N=0$, then so is $p$.
Thus just requiring chirality imposes the above non-trivial restriction on the
possible choices of $R^{2}$.

The other non-trivial condition we obtain is from requiring the existence of
spinorial states in the theory.  This requirement is physically important
because we know
that the Hall system must (unlike superconductivity) be described
microscopically using particles with fermionic statistics.  For example,
properties of the Hall fluid like incompressibility follow only if the
fundamental excitations are fermionic.   Such a requirement translates into
requiring that the operator which rotates the whole system by $2\pi$ has $-1$
as one of its eigenvalues.  The corresponding eigenstate would then be
spinorial and would describe a fermionic excitation.

Now, the operator  that
generates spatial translations along the boundary of the disc is
\be
\hat{P} = :\int _{0}^{2\pi}\Pi \phi ':=\hat{P}_{0} +\hat{P}_{oscillators},
\label{2.17}
\ee
where the double dots refer to normal ordering with respect to the oscillator
modes while the subscripts $0$ and $oscillators$ indicate the splitting of the
operator appearing in the integrand into contributions from the zero
and  oscillatory modes respectively.  The zero modes here refer to all modes
which are not oscillator modes and hence include the spatially constant as well
as the winding (soliton) modes.

The operator that translates by a distance $2\pi$ (along the edge) is the same
as the operator
which physically rotates the system by $2\pi$.  This operator is
\begin{eqnarray}
e^{i2\pi\hat{P}} & = &
e^{i2\pi\hat{P}_{0}}e^{i2\pi\hat{P}_{oscillators}}\nonumber\\
& = & e^{i2\pi\hat{P}_{0}}\label{2.18}
\end{eqnarray}
The last equality above is because the second factor acts as identity on all
states.  The easiest way to see this is by noticing that
any eigenstate of the number operator (in the Fock space of the
oscillator modes) is an eigenstate of $\hat{P}_{oscillator}$ with an
integer as the eigenvalue.  Thus $e^{i2\pi\hat{P}_{oscillator}}$ has 1 as
eigenvalue on all such states.  Since these states form a basis, we have the
above result that this factor acts as the identity operator.

Therefore,
\[ e^{i2\pi\hat{P}}  =  e^{i2\pi\hat{P}_{0}} = e^{ip
\int dx \phi '(x)}\]
which has the eigenvalues
\begin{eqnarray}
&  & e^{i2\pi (m+\alpha )N}.\label{2.19}
\end{eqnarray}
Since we require the above operator to have $-1$ as one of its eigenvalues, it
follows that there must exist some $m,\alpha$ and $N$ such that
\be
2\pi (m+\alpha )N =\pi \times \mbox{ odd integer}.\label{2.20}
\ee
If we limit the possibilities of $\alpha$ to just 0 or $\frac{1}{2}$
(corresponding to the wave function in (\ref{2.7}) being periodic
or antiperiodic), then the above condition can be rewritten as
\begin{eqnarray}
MN & =&\mbox{ odd integer},\nonumber\\
M&:=&2(m+\alpha ),\nonumber\\
M,N&\in &{\bf Z}.\label{2.21}
\end{eqnarray}
  It follows from (\ref{2.21}) that $M$ and $N$ are both odd
integers (and hence nonzero).

The condition (\ref{2.20}) can also be obtained by requiring
the existence of anticommuting vertex operators (VO's) that create
chiral fermions from the vacuum as we shall see in Section 2.2.

  Thus for $\alpha =0$ or 1/2, what we have in summary, for the case of one
chiral boson, is
\begin{eqnarray}
\mbox{Spectrum of $p$ $\equiv$ Spec }(p) &=& \{ \frac{M}{2}\} ,\nonumber\\
\int dx\phi '(x) & = & 2\pi N,\nonumber\\
R^{2} & = & \frac{M}{N}>0,\label{2.22}  \\
M,N&\in & 2{\bf Z}+1. \nonumber
\end{eqnarray}

This restriction on the possible
choices of $R^{2}$ along with (\ref{2.298}) leads to the result:
\be
\sigma _{H}=\frac{e^{2}}{2\pi}\frac{M}{N}.\label{2.30}
\ee
Thus the Hall conductivity here is quantized in fractions with not just odd
denominators, but also odd numerators.

It is important to emphasize here that
had we not modified the current (\ref{2.295}) to (\ref{2.296}) to take into
account the
contribution from the interior of the disc, we would have obtained a value for
the Hall conductivity which would have been physically {\it wrong} because it
would have predicted the existence of only even denominators (contrary to what
is experimentally observed).

If we have many independent scalar
fields at the boundary,  $\sigma _{H}$
will be the total of such fractions.
Thus $\sigma _{H}$ can take any fractional value with odd denominators
(there being no
condition that the numerator is also odd now).  To arrive at this
result, we of course also need the assumption that the charges $q_{i}$ of the
various
scalar fields are all integer multiples of the electronic charge $e$.

One further point about (\ref{2.30}) deserves mention.  This has to do with the
perplexing fact that we obtain fractions with $M\neq 1$ (in fact equal to any
odd number) even with a single chiral boson.  Normally, one obtains only
fractions with one in the numerator with a single chiral boson.  The reason for
this is that the states permitted in our theory contain all states of the form
$|\pm M_{0},\pm N_{0}\rangle ,\; |\pm 2M_{0},\pm 2N_{0}\rangle ,\ldots$ where
$R^{2}=\frac{M_{0}}{N_{0}}$ with $M_{0},N_{0}$ being co-prime (and odd). [The
arguments in these states indicate the eigenvalues of $2p$ and the winding
number.  They are vacua for the oscillator modes.]

{}From the definition (\ref{2.296}) of the current ${\cal J}^{\mu}$, we see
that
the charge defined as $\int dx {\cal J}^{0}$
has the spectrum of values
\be
\frac{e}{2}(M+R^{2}N)-\frac{e^{2}M}{2\pi N}\int dx
A_{1}=eM-\frac{e^{2}M}{2\pi N}\int dx A_{1}. \label{chaspe}
\ee
This equation
shows that the absolute value for the charge of the vacuum itself (which is the
 state $|0,0\rangle$ with no oscillator excitations) could be
fractional since $\int dx A_{1}$ could be an arbitrary real number.  In this
case,
the entire spectrum for the charge is also fractional, though integer spaced
(in units of $e$). Also the fundamental fermions of our theory are seen to have
 charges $=\pm eM_{0}$ with respect to the vacuum.

If we require the electron to exist as one of the
excitations of the theory, then we are forced to have $M_{0}=1$.
In that case the Hall conductivity is of the form
$\frac{e^{2}}{2\pi}\frac{1}{N_{0}}$.
\\
{\bf Duality and the QHE}
\nopagebreak
\par
We now come to the next result of this paper.  This has to do with
the duality ``symmetry'' (``$T$-duality'')that exists for the action
(\ref{2.1}) under the
transformation $R\rightarrow
\frac{1}{R}$ \cite{kik,bala}.  That this is a ``symmetry'' for
the Hamiltonian (in the sense of leaving its spectrum invariant) can be seen
most simply by looking at its explicit expression.  Since only the ``zero''
modes
(the spatially constant and solitonic modes)
of the theory contain information about $R$ (it being possible to
scale the non-zero modes
freely), we need focus attention also only on these zero modes.
The Hamiltonian restricted to these modes is
\begin{eqnarray}
H_{0} & = & \frac{R^{2}}{8\pi}[\frac{8\pi p^{2}}{R^{4}} + (\frac{\int dx
\phi '(x)}{2\pi})^{2}2\pi ] \label{hzero}
\end{eqnarray}
For the states $|M,N\rangle$ introduced above, we have
\begin{eqnarray}
p|M,N\rangle &=&\frac{M}{2}|M,N\rangle ,\nonumber\\
\frac{1}{2\pi}\int dx \phi '(x) |M,N\rangle &=& N|M,N\rangle, \label{modzero}
\end{eqnarray}
and so
\begin{eqnarray}
H_{0}|M,N\rangle & = &\frac{1}{4}(\frac{M^{2}}{R^{2}}
+N^{2}R^{2})|M,N\rangle .\label{2.315}
\end{eqnarray}
Hence the spectrum of $H_{0}$  remains invariant under $R\rightarrow
\frac{1}{R}$ [the eigenvalue of $H_{0}$ remaining unaltered if
$|M,N\rangle$ is replaced by $|N,M\rangle$ after this transformation].

 By itself, the above result is common knowledge.  The interest
in it for
us is for the following reasons.  $R^{2}$ is allowed to have only
certain rational values. Up to factors, it is also the Hall
conductivity for the theory on the disc whose edge excitations are described by
the action (\ref{2.1}).  As a consequence, $R\rightarrow \frac{1}{R}$
takes us from a Hall conductivity $\sigma _{H} = \frac{e^{2}}{2\pi}\frac{M}{N}$
to a new Hall conductivity $\sigma _{H}'=
\frac{e^{2}}{2\pi}\frac{N}{M}$.  This gives a novel way of relating Hall
conductivities at two fractions which are inverses of each other.  In
particular, it relates the integer Hall effect to the fractional Hall effect at
one
of the Laughlin fractions.

Physically, what this means is the following: the two theories described by $R$
and
$\frac{1}{R}$ are identical so long as they are free.  [In fact they
 are unitarily related in the sense that there exists a unitary
transformation on the operators which transforms the Hamiltonian $H_{0}$ into
another of the same form with $R$ replaced by $\frac{1}{R}$.]  However, these
theories are no longer identical after gauging since electromagnetism breaks
the symmetry.  The
Hall conductivities they give correspond to reciprocal  filling
fractions.   This is the
interesting result that we explore further in the subsequent sections.
Note that this result, in particular, means that the {\em spectra of edge
excitations at these two filling fractions are identical.}

{\bf 2.2. Mode expansion of the Chiral Boson}
\nopagebreak
\par
Until now, we have been working with a Lagrangian description of a scalar field
theory having both left and
right moving modes and then imposing chirality as a constraint on the system.
We could instead have followed the procedure (standard in conformal field
theory) of writing the scalar field itself as the sum of a left moving
piece $\phi _{l}$ and
a right moving piece $\phi _{r}$, each of which is separately
compactified on a circle:
\be
\phi (x,t)=\phi _{l}(x,t) +\phi _{r}(x,t).\label{2.33}
\ee
The $\phi _{l}$ and $\phi _{r}$ here depend respectively only on $x_{+}(=t+x)$
and $x_{-}(=t-x)$ so that the $t$-dependence of these fields will sometimes not
be specified explicitly.  In this approach, we impose chirality by the
constraints
\be
\frac{\partial}{\partial x_{-}}\phi _{r}^{(0)}(x,t)|\rangle
=\frac{\partial}{\partial x_{-}}\phi _{r}^{(+)}(x,t)|\rangle =0, \label{2.32}
\ee
on the physical states, where the superscripts $0$ and $+$ refer to
the  zero and positive
frequency components of $\frac{\partial}{\partial x_{-}}\phi _{r}(x,t)$.

The mode expansions in the original theory with the action (\ref{2.1})
are
\begin{eqnarray}
\phi (x,t) &=& q +Nx +\frac{2}{R^{2}}pt+\frac{1}{R}\sum _{n>0}[\frac{a_{n}}
{\sqrt{n}}e^{-inx_{+}}+
\frac{a_{n}^{\dagger}}{\sqrt{n}}e^{inx_{+}} + \frac{b_{n}}{\sqrt{n}}e^
{-inx_{-}}+\frac{b_{n}^{\dagger}}{\sqrt{n}}e^{inx_{-}}],\nonumber \\
\Pi (x,t) &=& \frac{p}{2\pi} -\frac{iR}{4\pi}\sum _{n>0}
[\sqrt{n}a_{n}e^{-inx_{+}}-
\sqrt{n}a_{n}^{\dagger}e^{inx_{+}} + \sqrt{n}b_{n}e^{-inx_{-}}-
\sqrt{n}b_{n}^{\dagger}e^{inx_{-}}].\label{2.35}
\end{eqnarray}
  The commutators of the operators in these expansions
follow from those of $\phi$ and $\Pi$ [(\ref{2.31})], the non-vanishing
commutators being
\begin{eqnarray}
&&[q,p]=i,\nonumber\\
&&{[}a_{n},a_{m}^{\dagger}{]}=\delta
_{nm}={[}b_{n},b_{m}^{\dagger}{]}.\label{new 10}
\end{eqnarray}

We write the mode expansions of $\phi _{l}$ and $\phi _{r}$ as
\begin{eqnarray}
\phi _{l}(x,t) &=& q_{l} +\frac{p_{l}}{R^{2}}x_{+} + \frac{1}{R}\sum
_{n>0}[\frac{a_{n}}{\sqrt{n}}e^{-inx_{+}}+\frac{a_{n}^{\dagger}}{\sqrt{
n}}e^{inx_{+}}],\nonumber\\
\phi _{r}(x,t) &=& q_{r} +\frac{p_{r}}{R^{2}}x_{-} + \frac{1}{R}\sum
_{n>0}[\frac{b_{n}}{\sqrt{n}}e^{-inx_{-}}+\frac{b_{n}^{\dagger}}{\sqrt{
n}}e^{inx_{-}}].\label{2.34}
\end{eqnarray}
Here, in view of (\ref{2.33}), we have
\begin{eqnarray}
q&=& q_{l} +q_{r}, \nonumber\\
p&=& \frac{p_{l} +p_{r}}{2},\nonumber\\
N&=& \frac{p_{l} -p_{r}}{R^{2}}.\label{2.36}
\end{eqnarray}
We now impose the following commutation relation on $q_{l,r}$ and
$p_{l,r}$ consistently with (\ref{new 10}):
\begin{eqnarray}
&&[q_{l},p_{l}] =i=[q_{r},p_{r}],\nonumber\\
&&\mbox{ All other commutators of these operators } =0.\label{2.38}
\end{eqnarray}

Note that $q_{l,r}$ and $p_{l,r}$ are not completely determined by $q$, $p$ and
$N$.  Rather we have introduced the additional operator
\be
\theta = \frac{R^{2}}{2}(q_{l}-q_{r})\label{2.37}
\ee
conjugate to $N$,
\be
[\theta ,N]=i \label{new 11}
\ee
in writing the expansions (\ref{2.34}).

We now look at the consequence of the chirality condition (\ref{2.32})
on $|\rangle $.
Because of (\ref{2.32}) and (\ref{2.34}) we have the result
\be
p_{r}|\rangle =b_{n}|\rangle =0.\label{2.40}
\ee
Since $p_{r}=p-\frac{R^{2}}{2}N$, we find (\ref{2.11}) again for $N\neq 0$:
\be
R^{2} =\frac{2p}{N}=\frac{p_{l}}{N}\mbox{ because }p_{r}=0.\label{2.41}
\ee
In this equation $p,p_{l},p_{r}$ and $N$ are to be interpreted as the
eigenvalues of
the corresponding operators on any physical state where they are all
diagonal.
  We shall use such a
convention whenever convenient to economise on symbols.  It will be
clear from the context if we are refering to operators or their
eigenvalues.

The following consequence of the chirality constraint is to be noted.
Since $2\times$ [any eigenvalue of $p$] is odd on our spinorial
states by (\ref{2.22}), (\ref{2.36}) and
(\ref{2.40}) show that
\be
\label{peru}
 \mbox{Spec } p_{l}\equiv \{ \mbox{ the eigenvalues of }p_{l}\} =2{\bf Z}+1
\ee
 on the physical subspace of spinorial states.

Let $|0\rangle $ denote the vacuum state.  It is annihilated by
$p_{l,r}$ and all annihilation operators.  We next construct the
vertex operators (VO's) that create solitons from $|0\rangle$.  Since
we are in the left-handed chiral subspace, we consider only the VO's
constructed out of $\phi _{l}$.  We thus consider the Fubini-Veneziano
VO \cite{fv}
\be
V(x_{+})=:e^{iM\phi _{l}(x_{+})}:,\label{2.42}
\ee
where the normal ordering symbol indicates that the $q_{l}$'s and
$a_{n}^{\dagger}$'s stand to the left of the $p_{l}$'s and $a_{n}$'s. [The
$t$-dependence of the operators here are given by their $x_{+}$-dependences.]

Now since $[p_{l},V(x_{+})] =MV(x_{+})$, $M$ has to be an odd integer if
$V(x_{+})|0\rangle$ is to be a spinorial (physical) state.  Similarly since
\be
[N,V(x_{+})] =[\frac{p_{l}}{R^{2}}, V(x_{+})] =\frac{M}{
R^{2}}V(x_{+}),\label{2.43}
\ee
 $\frac{M}{R^{2}}$ is the soliton number of $V(x_{+})|0\rangle $ and so
has to be an odd integer too.  By the way we see that $p_{l}$ and $N$
for the state
$V(x_{+})|0\rangle$ automatically satisfy the chirality constraint
(\ref{2.40}).

The charge of the  state $V(x_{+}')|0\rangle$ (with respect to the vacuum; see
(\ref{chaspe})) is calculated by considering the following commutation relation
at equal times: [All quantities below are at equal times and hence their
$t$-dependences are not shown.]
\begin{eqnarray}
[{\cal J}_{0}(x), V(x')] &=& \frac{eR^{2}}{4\pi}[D_{0}\phi (x) +D_{1}\phi (x)
,V(x')]=\frac{eR^{2}}{4\pi}[\dot{\phi}(x) +\phi '(x), V(x')]\nonumber\\
&=& e[\Pi (x) +\frac{R^{2}}{4\pi}\phi '(x),V(x')]\nonumber\\
&=& Me\delta (x-x')V(x').\label{2.44}
\end{eqnarray}
Thus the charge of $V(x'_{+})|0\rangle $ is $Me$ and it is located at the point
$x'$.

A calculation using methods standard in string and conformal field
theories shows that the requirement
\be
{[} V(x),V(x')] _{+} =0 \mbox{ for }x\neq x' \label{2.45}
\ee
at equal times forces $p_{l}N$ (and hence $p_{l}$ and $N$) to be odd, a
condition also required if
$V(x)|0\rangle$ is to be spinorial.  Thus the spinorial excitation
described by $V(x)|0\rangle$ obeys Fermi statistics.

The crucial advantage of the approach followed in this sub-section is that we
can now calculate the `$n$-particle' wave function of the minimum energy state
for the theory that lives at the boundary.  The minimum energy
state of $n$ particles, each with winding number $\frac{M}{R^{2}}$,
does not have any oscillatory excitations and is hence
\be
|(nMe)\rangle =e^{inMq_{l}}|0\rangle .\label{2.47}
\ee
Thus the $n$-particle wave function is (see also \cite{fub})
\be
\Psi (x_{1},x_{2},\ldots x_{n}) =\langle (nMe)|V(x_{1})V(x_{2})\ldots
V(x_{n})|0\rangle ^{*}.\label{2.48}
\ee
This expression gives, on using the mode expansions (\ref{2.34}) and the
CR's (\ref{new 10},\ref{2.38}),
\be
\Psi (x_{1},x_{2},\ldots x_{n})=\prod _{0<i<j}^{n}
(e^{-ix_{i}}-e^{-ix_{j}})^{\frac{M^{2}}{R^{2}}}=\prod _{0<i<j}^{n}
(e^{-ix_{i}}-e^{-ix_{j}})^{MN} .\label{2.49}
\ee

Now there is an intimate relationship between (\ref{2.49}) and Laughlin's
wave function for filling fraction $R^{2}=1/|N_{0}|$ when
$V(x)|0\rangle$ has charge $-e$ ( so that $M_{0}=-1$).  This was first
pointed out by Fubini, and Fubini and Lutken \cite{fub} and
studied further by many authors \cite{froh,wen,dim,kar,cap,nrea,cris}.  The
significance of this
relationship is particularly clear in our approach. Thus let $z=\xi
+i\eta$ be the complex coordinate on the disk $D$ of unit radius so
that $|z|=1$ at its boundary $\partial D$.  The Laughlin wave function
for $n$ particles of charge $-e$ for filling fraction $1/|N_{0}|$ in the
``symmetric'' gauge is
\be
\chi (\xi _{1}, \eta _{1}, \xi _{2}, \eta _{2},\ldots \xi _{n}, \eta
_{n})\equiv \chi (\{ \xi _{i}, \eta _{i}\} ) =e^{-\frac{1}{4 l^{2}}\sum
_{i=1}^{n}|z_{i}|^{2} }\prod
_{i<j}(\bar{z_{i}}-\bar{z_{j}})^{|N_{0}|},\label{Lwfn}
\ee
where $l$ is the cyclotron radius of the electron in the given magnetic
field and
$z_{i}=\xi _{i} +i\eta _{i}$ is the coordinate of the $i$th
particle on the disk.  Equation (\ref{Lwfn}) shows that
$e^{\frac{1}{4l^{2}}\sum
_{i=1}^{N} |z_{i}|^{2}}\chi (\{ \xi _{i}, \eta _{i} \} ) $ is an
anti-holomorphic function in all $z_{i}$ on $D$.  [Here we have chosen the
measure defining the scalar product of wave functions to be the
standard Lebesgue measure $\prod _{i}d\xi _{i}d\eta _{i}$.]

Now suppose we are given this information, namely that $e^{\frac{1}{4l^{2}}\sum
_{i=1}^{N} |z_{i}|^{2}}\chi (\{ \xi _{i}, \eta _{i} \} ) $ is an
anti-holomorphic function in all $z_{i}$ on $D$.  \underline{Then its
values when all $z_{i}$ are restricted to $\partial D$} \underline{completely
determines it everywhere on $D$ by analytic continuation.}

For the ground state wave function (\ref{2.49}) at the boundary, this
analytic continuation gives precisely the Laughlin wave function
(\ref{Lwfn}).  We can also of course construct excited state wave
functions at the edge and the coresponding analytically continued wave
functions.  It is in this manner that the Laughlin wave functions get
related to correlation functions of the conformal field theory at the
edge.

It is important to note that for this $M=-1$ case, we can only get the
filling fractions $1,1/3, 1/5,\ldots$ .

For pedagogical reasons, let us also record the wave function of the state
$a_{k}^{\dagger}|nMe\rangle $ which has an oscillator mode as well
excited.  The wave function of this state at $\partial
D$ is
\be
\tilde{\Psi} (x_{1},x_{2},\ldots ,x_{n})=\langle
(nMe)|a_{k}V(x_{1})V(x_{2})\ldots V(x_{n})|0\rangle ^{*}
\label{ewfn}
\ee
Since
\be
V(x)^{-1}a_{k}V(x) = a_{k} +i\frac{M}{\sqrt{k}R}e^{ikx}, \label{annih}
\ee
we readily find that
\be
\tilde{\Psi} (x_{1},x_{2},\ldots ,x_{n})=-i\frac{M}{\sqrt{k}R}(\sum
_{i=1}^{n}e^{-ikx_{i}}) \Psi (x_{1},x_{2},\ldots ,x_{n}) \label{elwf}
\ee
which continues antianalytically to the wave function
\be
\tilde{\Psi} (\bar{z_{1}},\bar{z_{2}},\ldots
,\bar{z_{n}})=-i\frac{M}{\sqrt{k}R}(\sum _{i}\bar{z_{i}}^{k})\Psi
(\bar{z_{1}},\bar{z_{2}},\ldots ,\bar{z_{n}}) \label{eLwf}
\ee
on $D$.

In these calculations, if from the beginning we had retained only
right-moving chiral modes at the edge, and imposed the hypothesis
that wave functions on $D$ are \underline{holomorphic} functions on $D$
up to the factor $e^{-\frac{1}{4 l^{2}}\sum |z_{i}|^{2}}$, the result would
be holomorphic versions of (\ref{2.49}, \ref{eLwf}).  If \underline{both}
left- and right- moving modes are excited at $\partial D$, the wave function at
$\partial D$ factorizes into a product of wave functions for left- and
right- moving pieces.  On requiring the wave function on $D$ to be the
product of the anti-holomorphic continuation of the former and
holomorphic continuation of the latter and the overall factor $\exp
(-\frac{1}{4 l^{2}} \sum |z_{i}|^{2})$, we find as usual that it is uniquely
determined.  The coefficient of $\exp (-\frac{1}{4 l^{2}}\sum |z_{i}|^{2})$
in this wave function of course contains both $z$'s and $\bar{z}$'s.

This approach to wave functions on $D$ from those on $\partial D$,
being quite general, will also work when the particle considered is
not an electron but some other quasiparticle (so that the charge $ke$ of
its state $:e^{ik\phi _{l}(x_{+})}:|0\rangle$ is not $-e$).  For example, we
can constrain
any allowed excitation [``quasiparticle''] by requiring that the wave function
of the
two-particle system consisting of the electron and the quasiparticle
is single-valued on $D$ (in addition to fulfilling the condition of
anti-holomorphicity outlined above).  Assuming that its restriction
to $\partial D$ is the ground state wave function of edge dynamics,
this edge wave function is seen to be
\be
\langle (k-1)e|:e^{-i\phi _{l}(x_{1})}::e^{ik\phi _{l}(x_{2})}: |0\rangle ^{*}=
(e^{-ix_{1}}-e^{-ix_{2}})^{-k|N_{0}|}.\label{quasi} \ee
The wave function on $D$ is thus
\be
e^{-\frac{1}{4l^{2}}(|z_{1}|^{2}+|z_{2}|^{2})}(\bar{z}_{1}-\bar{z}_{2})^
{-k|N_{0}|} \label{quasiw}
\ee
It is devoid of poles and branch cuts (and so finite and single-valued) only if
\be
k|N_{0}|\in {\bf Z}^{-} \mbox{ or } k\in \frac{{\bf Z}^{-}}{|N_{0}|}
\label{quacon}
\ee
where $R^{2}=1/|N_{0}|$.  This reproduces a well-known result in
Laughlin's theory \cite{laugh}.

The quasihole excitation of Laughlin is obtained by the choice
$k=-1/|N_{0}|$ and has previously been discussed by Fubini and Lutken
\cite{fub}.
A treatment of quasiparticle excitations is also developed in their work.

\sxn{$O(d,d;{\bf Z})$ and FQHE for $d$ Chiral Bosons}

{\bf 3.1. Chiral Constraint on Left-Right Symmetric Bosons}
\nopagebreak
\par
In this section, we generalise the action (\ref{2.1}) to one with many scalar
fields.  The motivation for doing this is that FQHE at fractions involving
numerators not equal to 1 is believed to be properly described by excitations
around many filled Landau levels.  Since every filled Landau level should give
rise to one chiral field at the edge, one may expect that many chiral fields
are required to
describe these excitations.  We shall see later however that there are
difficulties in obtaining the wave functions of more than one filled Landau
level in this approach.  One other reason for several scalar fields could be
that the different fields correspond to the edge fields of the different
excitations that
are permitted in the interior (namely, the statistical gauge fields and the
successive vorticial excitations that appear in the usual
hierarchical approaches).  In this approach, the first of the many scalar
fields alone describes the electronic edge current while the remaining fields
correspond to the edge currents of the statistical and vorticial fields.  It is
this approach
which also gives wave functions that can be interpreted as the
ground state wave functions of the appropriate filling fractions.  The
same is not possible with the approach using many filled Landau levels as will
be discussed below.

We will discuss the above-mentioned relation of bulk and edge excitations in
some detail in a work under preparation \cite{dua2}.

With many scalar fields at the edge, the action is
\be
A_{0}' = \frac{1}{8\pi}G_{ij}\int dt\int _{0}^{2\pi}dx\; \partial _{\mu}\phi
^{i}\partial ^{\mu}\phi ^{j} \label{3.1}
\ee
where $\phi ^{i}(x)$ is identified with $\phi ^{i}(x) +2\pi$.  $G$ here is
an invertible positive definite symmetric matrix with constant elements.

As in
string theories \cite{nar,gre}, we can also add  a topological term to the
above which
leaves the equations of motion unaffected to obtain the action
\be
A_{0} = \frac{1}{8\pi} G_{ij}\int d^{2}x\;\partial _{\mu}\phi ^{i}\partial ^
{\mu}\phi ^{j}+\frac{1}{8\pi}B_{ij}\int d^{2}x\; \partial _{\mu}\phi ^{i}
\partial _{\nu}\phi ^{j}\epsilon ^{\mu\nu} \label{3.2}
\ee
Here $B$ is a constant antisymmetric matrix:
\be
B_{ij}=-B_{ji} \label{Bij} .
\ee

Chirality for this theory can be enforced by imposing the condition
\be
\partial _{-}\phi ^{i}\equiv \frac{1}{2}(\dot{\phi}^{i}-\phi
^{'i})=0\label{3.3}
\ee
which eliminates the right-moving modes.  As before, we should check that such
a constraint is compatible with the
equations of motion of the action (\ref{3.2}).  Since the latter are
\be
\partial _{+}\partial _{-}\phi ^{i}=0,\label{3.31}
\ee
the two are mutually compatible.

However, as previously, this is no longer true when we gauge (\ref{3.2}):
  the gauge invariant chirality constraint and the equations of motion from the
gauged version of (\ref{3.2}) are not consistent as we will see below.

The gauged action is
\begin{eqnarray}
S_{0}(\phi ^{i},A)& =& \frac{1}{8\pi}G_{ij}\int d^{2}x\; D_{\mu}\phi ^{i}
D^{\mu}\phi ^{j}
+\frac{1}{8\pi}B_{ij}\int d^{2}x\; \partial _{\mu}\phi ^{i}\partial _{\nu}\phi
^{j}\epsilon ^{\mu\nu},\nonumber\\
D_{\mu}\phi ^{i}&\equiv &\partial _{\mu}\phi ^{i}-Q^{i}A_{\mu},\label{3.15}
\end{eqnarray}
where $Q^{i}$ is the charge associated with $\phi ^{i}$.
The last term in (\ref{3.2}) is not gauged here because it is a topological
term
and is gauge invariant.  We do not therefore alter it in the process of
gauging.

Imposition of the gauge invariant constraint
\be
D_{-}\phi ^{i}=(\partial _{-}\phi ^{i} -Q^{i}A_{-})=0
\label{3.16}
\ee
once again leads to an inconsistency with the equations of
motion
\be
\partial _{\mu} D^{\mu}\phi ^{i} =0 \label{eom}
\ee
of (\ref{3.15}) because
\begin{eqnarray}
0=\partial _{+}(D_{-}\phi ^{i}) & = & \frac{1}{4}(\partial _{\mu}(D^{\mu}\phi
^{i})+ Q^{i}E)\nonumber\\
\Rightarrow \partial _{\mu}(D^{\mu}\phi ^{i}) & = &
-Q^{i}E.\label{3.17}
\end{eqnarray}
Thus imposition of chirality necessitates augmentation of the action
(\ref{3.15}) by an \underline{anomaly term} $\frac{Q^{i}G_{ij}}{4\pi}\int
d\phi ^{j}A$ for compatibility with the equations of motion:
\be
S[\phi ,A] = S_{0}[\phi ,A] +\frac{Q^{i}G_{ij}}{4\pi}\int d^{2}x\epsilon
^{\mu\nu}\partial _{\mu}\phi ^{j}A_{\nu}. \label{3.171}
\ee

   Just as before we first define a
current $j^{\mu}=-\frac{\delta S}{\delta
A_{\mu}}=\frac{Q^{i}G_{ij}}{4\pi}(D^{\mu}\phi ^{j}+\epsilon ^{\mu\nu}\partial
_{\nu}\phi ^{j})$ and then ``covariantise'' it
to get a current ${\cal J}^{\mu}$:
\be
{\cal J}^{\mu} = \frac{Q^{i}G_{ij}}{4\pi}
(D^{\mu}\phi ^{j}+\epsilon ^{\mu\nu}D_{\nu}\phi ^{j}).\label{3.172}
\ee
This current is not only gauge invariant, it also satisfies ${\cal J}_{-}=0$ as
an identity.  Thus this is to be interpreted as the physical Hall current as
opposed to $j^{\mu}$ which is neither gauge invariant nor chiral.  As in the
case with a single chiral boson, here too we can interpret this modification of
$j^{\mu}$ as
arising out of a Chern-Simons term in the bulk required for gauge invariance of
the total action.

Using (\ref{3.17}) and (\ref{3.172}), we get
\be
\partial _{\mu}{\cal J}^{\mu} =-\;\frac{Q^{i}G_{ij}Q^{j}}{2\pi}E.\label{3.173}
\ee
Thus
\be
\sigma _{H} = \frac{Q^{i}G_{ij}Q^{j}}{2\pi}.  \label{3.18}
\ee

Having obtained $\sigma _{H}$ in terms of $G_{ij}$, we can now see the
conditions that a semiclassical theory (where electromagnetism
is treated classically) will impose on $G_{ij}$ and $B_{ij}$.

Letting $\Pi
_{i}$ denote the momentum field canonically conjugate to $\phi ^{i}$, we get,
from (\ref{3.171}),
\be
\Pi _{i} = \frac{G_{ij}}{4\pi}D_{0}\phi ^{j} +\frac{B_{ij}}{4\pi}\phi
^{'j}+\frac{G_{ij}Q^{j}}{4\pi}A_{1}. \label{3.4}
\ee
We also have as a consequence of (\ref{3.16})
\be
\frac{G_{ij}}{4\pi}\int _{0}^{2\pi}dx (D_{0}\phi ^{j}-D_{1}\phi ^{j})=0   .
\label{3.41}
\ee
Therefore, in view of (\ref{3.4}),
\be
\int dx(\Pi _{i} -\frac{1}{4\pi}(G_{ij}+B_{ij})\phi ^{'j}) =0.\label{3.5}
\ee

Now as in (\ref{2.9}),
\begin{eqnarray}
\mbox{Spec }p_{i}&=&\{ m_{i}+\alpha _{i}:m_{i}\in {\bf Z}, 0\leq \alpha _{i}<
1\}
,\nonumber\\
p_{i}&:=& \int dx \Pi _{i},\label{spect}
\end{eqnarray}
$\alpha _{i}$'s being fixed in a given theory.  Also
\be
\int _{0}^{2\pi} dx \phi ^{'i}=2\pi N^{i},\;\; N^{i}\in {\bf Z}, \label{wind}
\ee
$N^{i}$ being winding numbers.

Equation (\ref{3.5}) leads to the condition
\be
(G_{ij}+B_{ij})N^{j}=2(m_{i}+\alpha _{i}) \label{3.6}
\ee
on $m_{i} +\alpha _{i}$ and $N^{i}$, which is analogous to (\ref{2.11}).

As in Section 2.1, we get further constraints by looking at the generator of
translations
\be
\hat{P} =:\int _{0}^{2\pi} \Pi _{i}\phi ^{'i}: =\hat{P}_{0} +
\hat{P}_{oscillators} \label{3.10}
\ee
(the subscripts here having the same meaning as they did in (\ref{2.17}))
from which we arrive at the operator $\hat{R}(2\pi )$ that rotates the system
by $2\pi$:
\begin{eqnarray}
\hat{R}(2\pi ) & = & e^{i2\pi \hat{P}} = e^{i2\pi \hat{P}_{0}}\nonumber\\
& = & e^{i2\pi (m_{i}+\alpha _{i})N^{i}}.\label{3.11}
\end{eqnarray}
Requiring that there exists a spinorial state $|\rangle $ in the theory
[so that
$\hat{R}(2\pi )|\rangle =-|\rangle$] gives
\be
2(m_{i}+\alpha _{i})N^{i}=\mbox{ odd integer}\label{3.12}
\ee
on that state.  Exactly, the same condition can be obtained by requiring
appropriate vertex operators to anticommute as we shall see later.

We adopt the choices 0 and  $\frac{1}{2}$ for $\alpha _{i}$.  We are thus
allowing the wave functions to be only either periodic or antiperiodic in the
spatially constant modes $\phi _{0}^{i}$ of $\phi ^{i}$.  With this assumption,
\be
 M_{i}=2(m_{i} +\alpha _{i})\in {\bf Z} \label{odmo}
\ee
and (\ref{3.12}) takes the form
\be
M_{i}N^{i} = \mbox{ odd integer on spinorial states.}\label{3.13}
\ee
Equations (\ref{spect}), (\ref{wind}), (\ref{3.6}) and (\ref{odmo}) can
now be rewritten as
\begin{eqnarray}
(G_{ij}+B_{ij})N^{j} & = &  M_{i},\label{3.14}\\
\mbox{Spec }p_{i} & = & \frac{1}{2}M_{i},\label{3.141}\\
\int dx\phi ^{'i} & = & 2\pi N^{i},\label{3.142}\\
M_{i},N^{i}&\in &{\bf Z}.\label{3.143}
\end{eqnarray}

{\bf $O(d,d;{\bf Z})$ and QHE}
\nopagebreak
\par
For the action (\ref{3.2}) also, we have a
generalisation \cite{dua} of the duality transformation of the action
(\ref{2.1}).  Again the easiest way
to see
this is by looking at the part $H_{0}$ of the Hamiltonian containing only the
zero modes since, as is well known, this is the only
part that has information about $G_{ij}$ and $B_{ij}$.  [As in Section 2, zero
modes here refer to all modes which are not oscillator modes.] The operator
$H_{0}$ is defined by
\be
H_{0}|M,N\rangle  = \frac{1}{4}\left( \begin{array}{cc}
                        M_{i} & N^{i} \end{array}\right)
\left( \begin{array}{cc}
              (G^{-1})^{ij} & -(G^{-1})^{ik}B_{kj} \\
              B_{ik}(G^{-1})^{kj} & G_{ij}-B_{ik}(G^{-1})^{kl}B_{lj}
             \end{array}\right) \left( \begin{array}{c}
                                      M_{j} \\
                                      N^{j} \end{array}\right)
|M,N\rangle \label{3.19}
\ee
where $M_{i}$ and $N^{i}$ satisfy the conditions (\ref{3.13}) and (\ref{3.14})
and $|M,N\rangle$ is the vacuum state for the oscillator modes.

We now show that the group $O(d,d;{\bf Z})$ acts on the matrix
\be
 {\cal M}=   \left( \begin{array}{cc}
              G^{-1} & -G^{-1}B \\
              BG^{-1} &\;\; G-BG^{-1}B
             \end{array}\right)   \label{Hnot}
\ee
and transforms the Hamiltonian to a new one with the same spectrum as the
original Hamiltonian.  Much of these results are known \cite{dua}.

Let
\be
\eta =\left( \begin{array}{cc}
                      0_{d\times d} &   1_{d\times d}\\
                      1_{d\times d} &   0_{d\times d}
             \end{array} \right) \label{metric}
\ee
where the subscripts refer to the dimensionality of the matrices.
It defines a metric with signature $(+,+,\ldots ; -,-,\ldots )$ with an equal
number of $+$'s and $-$'s.  The real $2d\times 2d$ matrices $g$ fulfilling
\be
g^{T}\eta g=\eta \label{o(d,d)}
\ee
thus constitute the group $O(d,d;R)\equiv O(d,d)$.

Now one can check that ${\cal M}^{T}\eta {\cal M}=\eta$ so that
\be
{\cal M}\in O(d,d).\label{calm}
\ee

Furthermore $g^{T}{\cal M}g={\cal M}'$ can also be written in the form
(\ref{Hnot}) with $G$ and $B$ replaced by $G'$ and $B'$ respectively.  The
sketch of the proof is as follows:

Firstly, we note that ${\cal M}'$, like ${\cal M}$, is symmetric and is an
element of $O(d,d)$.  Furthermore, it can be checked that the top left
$d\times d$ sub-matrix of ${\cal M}'$ is positive definite (and hence
non-singular) provided the same is true of the corresponding sub-matrix of
${\cal M}$ (as is the case because of our assumption that $G$ is a positive
definite symmetric matrix).
Using these properties, it can be shown that ${\cal M}'$ can be written
in the form (\ref{Hnot}) for some $G'$ and $B'$.

Thus $O(d,d)$ acts on matrices of the form ${\cal
M}$, or equally well, on the set of pairs $(G,B)$.  It thus transforms the
action $A_{0}$ to a new one with a changed $G$ and $B$.

But $O(d,d)$ does not act on $(M, N)$, as it does not preserve the
condition $M_{i},N^{i}\in {\bf Z}$.  It is only $O(d,d;{\bf Z})$ that does so.

Now $O(d,d;{\bf Z})$ also preserves the remaining conditions (\ref{3.13}) and
(\ref{3.14})
on $(M, N)$.  This is so because they are respectively equivalent to the
manifestly $O(d,d;{\bf Z})$ compatible conditions
\begin{eqnarray}
\frac{1}{2}\left( \begin{array}{cc} M & N \end{array} \right) \eta \left(
\begin{array}{c} M \\ N \end{array} \right) & \in & 2{\bf Z}+1 ,\label{m.n}\\
{\cal M}\left( \begin{array}{c} M\\ N \end{array} \right) &=& \eta \left(
\begin{array}{c} M\\ N \end{array} \right), \label{chir}
\end{eqnarray}
the action of $O(d,d;{\bf Z})$ being
\be
{\cal M} \rightarrow {\cal M}' =g^{T}{\cal M}g, \left( \begin{array}{c} M\\N
\end{array} \right) \rightarrow \left( \begin{array}{c} M'\\N' \end{array}
\right) =g^{-1}\left( \begin{array}{c} M\\N \end{array} \right) ,g\in
O(d,d;{\bf Z}).\label{graction}
\ee

It now follows readily that the spectrum of the Hamiltonian defined by $(G,B)$
is precisely the same as that of the one defined by $(G',B')$.  $O(d,d;{\bf
Z})$ hence
generalises the duality transformation of Section 2.  Note however that the set
of pairs $(M,N)$ fulfilling our conditions are not necessarily the same in
the two theories.

The summary of what we have seen in this section is the following.  By
considering a theory described by (\ref{3.1}) on the boundary of a disc and by
imposing chirality on the fields and currents, we see that the theory on the
disc has a
Hall conductivity given by $\sigma _{H}=\frac{1}{2\pi}Q^{i}G_{ij}Q^{j}$.
Moreover, the $G_{ij}$ and $B_{ij}$ have to satisfy certain relations
(\ref{3.13},\ref{3.14}) in order that they describe a chiral theory that
permits the existence of fermionic excitations.

Furthermore, there exists an $O(d,d;{\bf Z})$ group of transformations that
leaves the
spectrum of the Hamiltonian of (\ref{3.2}) invariant.  This happens because
these transformations on the matrix ${\cal M}$ and
the corresponding transformations on $(M,N)$ are compatible with the conditions
(\ref{3.13},\ref{3.14}).

Let $G'$ and $B'$ be the transform of $G$ and $B$ under an element of
$O(d,d;{\bf Z})$.  If $Q^{i}$'s are held fixed under this transformation, the
corresponding Hall conductivities are $\sigma
_{H}=\frac{1}{2\pi}Q^{i}G_{ij}Q^{j}$ and $\sigma '
_{H}=\frac{1}{2\pi}Q^{i}G'_{ij}Q^{j}$.  We thus see that we have a way of
relating $\sigma _{H}$ with $\sigma _{H}'$ using the group $O(d,d;{\bf Z})$.

In the next section, we will use this result to arrive at the hierarchy due
to Haldane \cite{hald} as well as the Jain fractions \cite{jain}.

{\bf 3.2. Mode expansions of Chiral Bosons}
\nopagebreak
\par
We begin with the mode expansions of $\phi ^{i}$ and $\Pi _{i}$ for the theory
described by the action (\ref{3.2}). [See also the definitions (\ref{spect},
\ref{wind})]:
\begin{eqnarray}
\phi ^{i}(x,t)&=& q^{i} +N^{i}x + (G^{-1})^{ij}(2p_{j}-B_{jk}N^{k})t + \sum
_{n>0} \frac{1}{\sqrt{n}}[a_{n}^{i}e^{-in(t+x)}+a_{n}^{i\dagger}e^{in(t+x)}]
+\nonumber\\ && \sum _{n>0}\frac{1}{\sqrt{n}}[b_{n}^{i}e^{-in(t-x)}+b_{n}^{i
\dagger}e^{in(t-x)}], \nonumber\\
\Pi _{i}(x,t) &=& \frac{p_{i}}{2\pi}-\frac{i}{4\pi}\sum _{n>0}\sqrt{n}
[(G+B)_{ij}
(a_{n}^{j}e^{-in(t+x)}-a_{n}^{j\dagger}e^{in(t+x)})]-\nonumber\\
&&\frac{i}{4\pi}\sum _{n>0}\sqrt{n}[(G-B)_{ij}(b_{n}^{j}
e^{-in(t-x)}-b_{n}^{j\dagger}e^{in(t-x)})] .\label{mexp}
\end{eqnarray}
The non-zero commutators between the operators occurring here are given by
\begin{eqnarray}
{[} q^{i},p_{j}{]}&=&i\delta ^{i}_{j},\nonumber\\
{[} a_{n}^{i},a_{m}^{j\dagger}{]}&=&\delta _{nm}(G^{-1})^{ij},\nonumber\\
{[} b_{n}^{i},b_{m}^{j\dagger}{]}&=&\delta _{nm}(G^{-1})^{ij}.\label{comrel}
\end{eqnarray}
The expansion of $\phi ^{i}(x,t)$ can also be rewritten in the slightly more
convenient form
\begin{eqnarray}
\phi ^{i}(x,t) &=& q^{i}+(G^{-1})^{ij}p_{lj}(t+x)+(G^{-1})^{ij}p_{rj}(t-x) +
\sum
_{n>0}\frac{1}{\sqrt{n}}[a_{n}^{i}e^{-in(t+x)}+a_{n}^{i\dagger}e^{in(t+x)}]
 +\nonumber\\
&&+\sum _{n>0} \frac{1}{\sqrt{n}}[b_{n}^{i}e^{-in(t-x)}+b_{n}^{i
\dagger}e^{in(t-x)}] ,\nonumber\\
&& p_{li} := p_{i} +\frac{1}{2}(G-B)_{ij}N^{j},\nonumber\\
&& p_{ri} := p_{i} -\frac{1}{2}(G+B)_{ij}N^{j}.\label{mexp2}
\end{eqnarray}

{}From the definition of $p_{l}$ and $p_{r}$ in (\ref{mexp2}), we have the
relations
\begin{eqnarray}
p&=& \frac{1}{2}[(G+B)G^{-1}p_{l} +(G-B)G^{-1}p_{r}], \label{plr}\\
N&=& G^{-1}(p_{l}-p_{r}).\label{nlr}
\end{eqnarray}

If we now introduce the operators $q_{l}^{i}$ and $q_{r}^{i}$ conjugate to
$p_{li}$ and
$p_{ri}$ respectively and commuting with all other operators, then we see that
$q_{l}^{i}+q_{r}^{i}$ is conjugate to
$p_{i}$ and moreover commutes with $N^{i}$ and the oscillator modes.  Thus
$q_{l}^{i}+q_{r}^{i}$ is the same as
$q^{i}$. We can now decompose $\phi ^{i}$ into left- and right-moving fields
$\phi _{l}^{i}$ and $\phi _{r}^{i}$ with the help of these operators thus:
\begin{eqnarray}
\phi _{l}^{i}(x_{+})&=&q_{l}^{i}+(G^{-1})^{ij}p_{lj}\; x_{+} +\sum
_{n>0}\frac{1}{\sqrt{n}}[a_{n}^{
i}e^{-inx_{+}}+\mbox{ h.\ c.} ],\nonumber\\
\phi _{r}^{i}(x_{-})&=&q_{r}^{i}+(G^{-1})^{ij}p_{rj}\; x_{-} +\sum
_{n>0}\frac{1}{\sqrt{n}}[b_{n}^{
i}e^{-inx_{-}}+\mbox{ h.\ c.} ],\nonumber\\
&&x_{+}:=t+x\;\; , \;\;x_{-}:=t-x. \label{mexp3}
\end{eqnarray}
Note that $\phi ^{i}(x,t)=\phi _{l}^{i}(x_{+})+\phi _{r}^{i}(x_{-})$ and that
$\phi _{l}$ and $\phi _{r}$ commute with each
other.

It should of course be verified that the above mode decomposition is preserved
in time.  For this to be true, we need
\begin{eqnarray}
{[} q_{l}^{i},H{]} &=&i(G^{-1})^{ij}p_{lj},\nonumber\\
{[} q_{r}^{i},H{]} &=&i(G^{-1})^{ij}p_{rj}.\label{evln}
\end{eqnarray}
These are indeed satisfied as can be checked by rewriting (\ref{3.19}) in terms
of the variables $p_{l}$ and $p_{r}$:
\be
H_{0}|p_{l},p_{r}\rangle =\frac{1}{2}\left( \begin{array}{cc} p_{li} &
p_{ri}\end{array} \right) \left( \begin{array}{cc} (G^{-1})^{ij}&0\\0&(G^{-1})
^{ij} \end{array} \right) \left( \begin{array}{c} p_{lj}\\p_{rj}\end{array}
\right) |p_{l},p_{r}\rangle.\label{hlr}
\ee

The chirality constraint can be imposed as previously by requiring
\be
\frac{\partial \phi _{r}^{i(0)}}{\partial x_{-}}|\rangle =\frac{\partial
\phi _{r}^{i(+)}}{\partial x_{-}}|\rangle =0, \label{chexp}
\ee
or
\be
p_{r}|\rangle =b_{n}|\rangle =0 \label{chexp2}
\ee
on any allowed physical state $|\rangle$.  The vanishing of $p_{r}$ on a
physical state $|\rangle$ with winding numbers $N^{i}$ and eigenvalues
$M_{i}$ for $2p_{i}$ gives back the result (\ref{3.14}):
\[ (G+B)_{ij}N^{j}=M_{i}. \]
The $M_{i}$'s here are constrained to be integers if we require the $2p_{i}$'s
to have an integral spectrum.  If we denote the eigenvalues of
$p_{li}$ by $M_{li}$, then we see from (\ref{plr}) and (\ref{chexp2}) that
\be
2p =(G+B)G^{-1}p_{l}, \mbox{ or } M=(G+B)G^{-1}M_{l}\label{chexp5}
\ee
Therefore, using also (\ref{3.14}), we get
\be
G_{ij}N^{j}=M_{li}. \label{chexp4}
\ee
It should be noted here that there is no requirement that the $M_{li}$'s are
integers.

The vertex operator which when acting on the vacuum creates a state with
winding numbers $N^{i}$ and $p_{li}=M_{li}$ (such that
(\ref{chexp4}) is fulfilled) is
\be
V(x_{+})=:e^{iM_{li}\phi _{l}^{i}(x_{+})}: \label{vertop}
\ee

As in the calculation in (\ref{2.44}), we can calculate the charge of the state
$V(x_{+})|0\rangle$ by commuting the charge density
operator ${\cal J}^{0}(x')$ (obtained from (\ref{3.172})) and $V(x)$ at equal
times. [Here by the charge of a state, we mean the deviation of its charge from
that of the vacuum.]  We find
that this charge is
\be
M_{li}Q^{i}\label{charge}
\ee
 and that it is localised at the point $x$.

The requirement
\be
{[ V(x), V(x')]} _{+}=0 \label{acvo}
\ee
at equal times leads to the condition
\be
M_{li}(G^{-1})^{ij}M_{lj}\in 2{\bf Z}+1.\label{condition}
\ee
Using
(\ref{chexp4}), (\ref{3.14}), and the facts that $G$ and $B$ are symmetric
and antisymmetric respectively, we now have the following set of identities:
\begin{eqnarray}
M_{li}(G^{-1})^{ij}M_{lj} &=& N^{i}M_{li} \nonumber\\
&=& N^{i}G_{ij}N^{j} \nonumber\\
&=& N^{i}(G+B)_{ij}N^{j} \nonumber\\
&=& N^{i}M_{i} . \label{identities}
\end{eqnarray}
Thus the condition under which (\ref{acvo}) is satisfied is same as the
condition written down in (\ref{3.13}) which was needed for the existence of
spinorial states.

The formalism developed in this sub-section can be used to find the
$n$-particle wave function of the minimum energy $n$-particle state at the
boundary.  Calculations analogous to those leading to (\ref{2.49})
give
\be
\Psi (\bar{z}_{1},\bar{z}_{2},\ldots \bar{z}_{n})=\prod
_{i,j=1;i<j}^{n}(\bar{z}_{i}-\bar{z}_{j})^{M_{lk}N^{k}}\label{3.32}
\ee
where $z_{i} =e^{ix_{i}}$, $x_{i}$ being coordinates on the circle. (The
exponent here can also be written as $M_{k}N^{k}$ since these two
expressions are equal owing to (\ref{identities}).)

We can of course think of obvious generalisations of the above wave function
when the particles are not all identical.  For example, we can consider the
state having $n_{1}$ particles with charge $M^{(1)}_{li}Q^{i}$, $n_{2}$
particles
with charge $M^{(2)}_{li}Q^{i}$ etc.\@  up to $n_{K}$ particles with charge
$M^{(K)}_{li}Q^{i}$.  For the particular case where
$M^{(r)}_{li}=\delta _{ri}$,  (chosen for the case of specificity) the
wave function of the minimum energy state at the boundary is
\be
\prod _{i,j=1;i<j}^{n_{1}}(\bar{z}^{(1)}_{i}-\bar{z}^{(1)}_{j})^{(G^{-1})^{11}}
\prod _{k,l;k<l}^{n_{2}}(\bar{z}^{(2)}_{k}-\bar{z}^{(2)}_{l})^{(G^{-1})^{22}}
\prod _{i=1}^{n_{1}}\prod
_{k=1}^{n_{2}}(\bar{z}^{(1)}_{i}-\bar{z}^{(2)}_{k})^{(G^{-1})^{12}}\ldots
,\label{3.33}
\ee
where $z^{(r)}_{i}$ is the coordinate of the $i$th particle of type $r$ (that
is, created by the operator $:e^{i\phi ^{r}_{l}(x_{i})}:$).

As in sub-section 2.2, we can analytically continue these wave functions into
the interior of the disc and they should then give the anti-holomorphic part of
the
ground state wave functions for the theory in the disc.  However, the
wave functions so obtained are not the multi-electron wave functions that one
would have got for filling fractions involving many filled Landau levels.  This
is because all these wave functions are anti-holomorphic by construction
whereas
we expect wave functions of electrons occupying higher Landau levels to contain
holomorphic pieces too.  We thus see that the present approach is inadequate
for describing a situation with many filled Landau levels.

However, the other interpretation of the scalar fields as representing the edge
currents of the different particles (electrons and vorticial charges/fluxons of
the successive hierarchies) makes sense because then we are still only in the
lowest Landau level.  It is therefore this approach that we shall focus on
while
computing the Hall conductivities.  Thus in particular, we shall associate the
first scalar field alone with an electronic charge and ascribe zero charges for
the remaining scalar fields when we later use (\ref{3.18}) to calculate
conductivities in Section 4.

Now, note that if these wave functions are to be well-defined
(singularity free) on the disk, then the entries of the
matrix $G^{-1}$ have to be integers:
\be
(G^{-1})\in {\bf Z} .\label{integral}
\ee
  Hence the wave function (\ref{3.33}) is
either symmetric or
antisymmetric under interchange of the order of the vertex operators creating
two different kinds of particles signifying that they are bosons or fermions.
In other words, we
are allowing the wave functions to be only either single-valued or
double-valued
under the transport of one particle around another.

The integrality condition (\ref{integral} is analogous to the integrality
condition $(R^{2})^{-1}\in {\bf Z}$ coming from (\ref{2.49}) for $M=1$.

Now we shall see in the
next section that an appropriate choice of the matrix $G+B$ reproduces the
Haldane
hierarchy under a sequence of $O(d,d;{\bf Z})$ transformations.

The matrix $(G+B)^{-1}$ appropriate for Haldane's hierarchy is given in
(\ref{4.1ti}).  We also note here that the corresponding wavefunctions given by
(\ref{3.33}) match the known hierarchical wave functions
\cite{hald,halp,nrea,cris}.

\sxn{$O(d,d;{\bf Z})$ and the Hierarchy}

	We now have a theory of $d$ chiral massless scalars describing the
edge states of a quantum Hall system.

Physical motivations for certain choices of $G$ come from the ``hierarchy''
scheme of Haldane \cite{hald} or from the approach of Jain \cite{jain}.
Perhaps the
simplest way to get restrictions on $G$ from the hierarchy schemes is to first
construct the Chern-Simons (CS) mean field theories for these schemes on the
disk.  The CS Lagrangian has previously been used by many authors
\cite{sri,read,zee,froh}
for the QHE.  In a paper under preparation \cite{dua2}, we will also discuss
its
use to get the Haldane hierarchy.  Now these CS models contain many connection
one-forms and a matrix $\tilde{G}$ coupling them.  There are also scalar fields
at the edge of the disk with an anomaly term containing $G$ as in
(\ref{3.171}).
Just as in the previous Sections, neither the edge nor the bulk actions are
separately gauge invariant, but one can show that their sum is, provided
$\tilde{G} =G$.  In this way the hierarchy schemes can restrict $G$.

We will elaborate on these considerations involving the hierarchy schemes in
\cite{dua2}.  Here we will content ourselves with discussing a method to obtain
the
continued fractions of the Haldane hierarchy for the Hall
conductivity $\frac{1}{2\pi}Q^{i}G_{ij}Q^{j}$ starting from the following
choice for $G^{-1}$:
\be
G^{-1}= \left[ \begin{array}{rrrrrr} m&0&0&.&.&.\\ 0&2p_{1}&0&0&0&.\\
0&0&2p_{2}&0&0&.\\ .&0&0&2p_{3}&0&.\\ .&.&.&.&.&. \end{array} \right]
\label{4.1}
\ee
Such a choice with
\be
m\in 2{\bf Z}+1,\;\;\; p\in {\bf Z} \label{wtni}
\ee
is the simplest choice for $G^{-1}$ compatible with physical requirements.  It
generalises the $R^{-2}$ for the single scalar field in (\ref{2.1}) to $d$
uncoupled
scalar fields.  At the same time, it ensures that the first field alone
permits fermionic vertex operators while the rest permit only bosonic vertex
operators.  Thus it permits only one electron field, namely the unique
fermionic
vertex operator, which is as we want (see the first paragraph of Section 3.1 in
this connection).

Let us first summarize our point of view.  The
$O(d,d;{\bf Z})$ transformations are ``symmetries'' of the ungauged theory, in
the sense that there is a well-defined mapping from the states of a theory
with given $G$ and $B$ to the states of a theory with transformed
$G$ and $B$ such that the spectrum of the Hamiltonian
remains invariant.
However once we gauge the theory, this ``symmetry'' is broken explicitly.
We first select the set of charges $Q^{i}$ as follows:
\be
Q^{1} = e,\;\; Q^{2} = Q^{3} =Q^{4} = \ldots = 0. \label{kya}
\ee
They are so chosen that the first field alone has the charge of an electron
while the
rest have zero charge.  If we now transform
$G$ and $B$ {\em keeping the $Q^{i}$ fixed}, we get a new value for the Hall
conductivity.  One can advance the hypothesis that the states thus
related are similar in their dynamical properties.  We would then be saying
that the ``strong'' interactions between the electrons respect this
$O(d,d;{\bf Z})$ ``symmetry'', but the (weak) external electromagnetic
interactions
break the ``symmetry''.  The symmetry being broken in a well-defined way
by a term in the action that transforms in a definite way, we get definite
predictions relating physical properties of one system with that of the
$O(d,d;{\bf Z})$ transformed system.  This is reminiscent of the isospin
symmetry
of strong interactions broken by electromagnetism. Thus, in our approach,  the
underlying dynamics of the Hall system obeys
the duality ``symmetry'' of the ungauged scalar theory describing the edge
states, but this is broken by the coupling with the (weak) external
electromagnetic field.

A choice of $G$ and $B$ determines the matrix ${\cal M}$ in (\ref{Hnot}), which
is an element of $O(d,d)$.  When we act on ${\cal M}$ with group
elements of $O(d,d;{\bf Z})$, we move along an orbit.  Different points
on an orbit correspond to systems with identical spectrum, but
with different response to external electromagnetic fields. We will
first illustrate the action of $O(d,d;{\bf Z})$ on the $G$ given by
(\ref{4.1}).
Later we
will discuss another choice that also generates a similar
hierarchy of filling fractions.

Let us begin by recalling the discussion of Section 2, where there is just one
scalar field.  The Hall conductivity in that model is given by (\ref{2.30}):
\[
\sigma _{H}=\frac{e^{2}}{2\pi}\frac{M}{N}\; ; \; M,N\in 2{\bf Z}+1 .\]

Now, the charge of the particle associated with $(M, N)$ is $Me$.  On
requiring that there is an electron in the theory, we see that our choice of
$R^{2}$ must admit a pair $(-1,-|N_{0}|)$ and hence also $(1,|N_{0}|)$
with $R^{2}=1/|N_{0}|$.  Thus we get filling fractions $1,1/3,1/5,\ldots $,
excitations characterized by $(M,N)=m(1,|N_{0}|)\;\; m\in {\bf Z}$, and
charges and winding numbers $me$ and $m|N_{0}|$ respectively.  The elementary
excitations have charges $\pm e$ and corresponding winding numbers $\pm
|N_{0}|$.

Since $R^{2}\rightarrow 1/R^{2}$ under duality, it maps $(M, N)$ to
$(N, M)$ or $m(1,|N_{0}|)$ to $m(|N_{0}|,1)$.  We can now get
integral filling factors while the elementary excitations have charges $\pm
|N_{0}|e$ and winding numbers $\pm 1$.

 Let us next turn to the two-scalar case.  The analogue of the constraint
$R^{2} = M/N$ is the chirality condition
\be
(G+B)_{ij} N^{j} = M_{i},    \; \; i=1,2 .  \label{4.27}
\ee
Let us consider
\be
\label{4.28}
B=0, G^{-1} = (G+B)^{-1}=\left( \begin{array}{cc}m & 0 \\ 0 &
 2 p \end{array} \right) ,
\ee
with $Q^{1}=e,\;\; Q^{2}=0$ (see (\ref{kya})).  On using (\ref{3.18}), we get
$\sigma _{H} = \frac{e^{2}}{2 \pi m}$.  The filling fraction $\nu$ is therefore
equal to $1/m$.  In this formula, $m$ has to be chosen to be odd since
otherwise there will be no solution for the two conditions
(\ref{3.13}) and (\ref{3.14}).  Thus, the filling fraction is forced to obey
the
odd denominator rule for the above choice of $G$ and $B$.

Let us define the transformations $E^{(v)}$ and $D$ by
\begin{eqnarray}
&&E^{(v)}: G\rightarrow G, \;\; E^{(v)}: B\rightarrow B+\epsilon ^{(v)}
,\nonumber\\
&&D: G+B \rightarrow (G+B)^{-1} . \label{haltra}
\end{eqnarray}
Here $\epsilon ^{(v)}$ is an antisymmetric matrix (having the same
dimensionality as $G$ or $B$) which
satisfies
\be
\epsilon ^{(v)}_{ij}=-\epsilon ^{(v)}_{ji}=\delta _{vi}\delta
_{v+1,j}-(i\leftrightarrow j). \label{epsiv}
\ee
  It can be checked that these two
transformations are both elements of $O(d,d;{\bf Z})$ and that the
corresponding
group elements (that act on ${\cal M}$ as in (\ref{graction})) are:
\begin{eqnarray}
E^{(v)} &\equiv &\left( \begin{array}{cc} 1 & -\epsilon ^{(v)}\\
                                  0 & 1 \end{array} \right) ,\nonumber\\
D &\equiv &\left( \begin{array}{cc} 0 & 1\\
                                  1 & 0 \end{array} \right) .\label{haltra1}
\end{eqnarray}
Here 1 and $\epsilon ^{(v)}$ denote unit and antisymmetric matrices of
the appropriate dimension.  These definitions will be used
for general $d\neq 2$ as well.

Let us now specialize to the case $d=2$.  Then $v$ has only one value and hence
will be omitted.  For this $d$, consider the sequence of transformations
\be
 (G+B)\stackrel{D}{\rightarrow}(G+B)^{-1}
 \equiv (G'+B')\stackrel{E}{\rightarrow} (G'+B'+\epsilon ) \equiv
(G''+B'')\stackrel{D}{\rightarrow} (G''+B'')^{-1}\equiv (G'''+B''').
\label{4.29}
\ee

After the first transformation, we get
\be
\label{4.30}
(G'+B') =
\left( \begin{array}{cc}m & 0 \\ 0 & 2 p \end{array} \right)  .
\ee
The value of the filling fraction $\nu$ is now $m$.  At the end of the second
transformation, we get
\be
\label{4.305}
(G''+B'') = \left( \begin{array}{cc} m&1\\ -1&2p \end{array} \right)
,\ee
so that the filling fraction $\nu$ is still equal to $m$.  At the end of
the third and final transformation, we get
\be
\label{4.31}
(G'''+B''') = \frac{1}{2pm+1}
 \left( \begin{array}{cc}2p & 1 \\-1 & m \end{array}\right)
,\ee
which gives the filling fraction
\be
 \nu = \frac{2p}{2pm+1}=\frac{1}{m+\frac{1}{2p}}\;\; .\label{hal2}
\ee

Thus we have for the action of the triplet $D*E*D$ on $\nu$,
\be
\frac{1}{m}\stackrel{D}{\rightarrow} m \stackrel{E}{\rightarrow} m
\stackrel{D}{\rightarrow} \frac{1}{m+\frac{1}{2p}}\;\; .\label{haltra2}
\ee
Calling the action $D*E^{(v)}*D$ as $\tilde{E}^{(v)}$, we see that we can write
a corresponding element of $O(d,d;{\bf Z})$ for it using (\ref{haltra1}):
\be
\tilde{E}^{(v)} \equiv \left( \begin{array}{cc} 1&0\\ -\epsilon ^{(v)}&1
\end{array} \right) \label{haltra3}
.\ee

Written in this form, the interpretation of $\tilde{E}^{(v)}$ is quite clear.
Namely, it is that transformation on ${\cal M}$ with the property that the
corresponding transformation on $(M,N)$ (see (\ref{graction})) leaves $M$
invariant and changes $N$ alone. The condition that $M$ and not $M_{l}$ is left
unchanged may
be some cause for concern at first sight.  Thus for example the invariance of
charge $M_{li}Q^{i}$ requires the invariance of $M_{l}$.  However, a little
thought shows that
it is precisely this feature of the above transformations that also permits
fractional charges for the transformed theory even though the original theory
had only integral charges (since $B=0$ and $M=M_{l}$ for the original theory).

	One can do such a transformation for the more general case
with $d$ scalar fields.  If we start with
\be
\label{4.34}
B=0,\;\; G^{-1}=(G+B)^{-1}= diag(m,2p_{1} , 2p_{2} ....)
\ee
We can perform the sequence of transformations
$\tilde{E}^{(d-1)}*\tilde{E}^{(d-2)}*\ldots \tilde{E}^{(1)}$, where
$\tilde{E}^{(v)}$ is defined in (\ref{haltra3}), to get a transformed
$(G+B)^{-1}$ which looks as follows:
\be
(G+B)^{-1}= \left[ \begin{array}{rrrrrr} m&1&0&.&.&.\\ -1&2p_{1}&1&0&0&.\\
0&-1&2p_{2}&1&0&.\\ .&0&-1&2p_{3}&1&.\\ .&.&.&.&.&. \end{array} \right]   .
\label{4.1t}
\ee
Once again it is easy to
see that we relate the state with $\nu = 1/m$ to the state with
\be
\nu = \frac{1}{m+ \frac{1}{2p_{1}+\frac{1}{2p_{2}+...}}}\;\; .\label{hienu}
\ee
 [See \cite{zee} for a
similar calculation performed with a matrix like the one in (\ref{4.1t})
except for the fact that the matrix in \cite{zee} is symmetric.] Thus we can
generate the entire hierarchy by means of these
generalized duality transformations on the Laughlin fractions \cite{laugh}!

While the orbit considered above does generate the required filling
fractions, the theories at its distinct points have, in general,
no vertex operators that create excitations with the charge of an
electron.  We shall see this in more detail in the following.

In the one scalar case, we have already seen that electron operators
exist when $R^{2} = \frac{1}{|N_{0}|}$ (but not when $R^{2} = |N_{0}|$), with
$|N_{0}|$ being odd.

Let us turn to the two scalar case.  The conditions that
have to be satisfied are:
\begin{eqnarray}
&&a)\;\;\;\;\;\; (G+B)_{ij}N^{j}=M_{i}  \label{eq:(1)} \\
\mbox{with $M,N$ being integers,}&&\nonumber\\
&&b)\;\;\;\;\;\; G_{ij}N^{j}=M_{li},\;\;\; M_{l1}=-1,      \label{eq:(2)}\\
\mbox{and}&&\nonumber\\
&&c)\;\;\;\;\;\; M,N\in {\bf Z};\;\;\; M_{i}N^{i}=2{\bf Z}+1 .  \label{eq:(3)}
\end{eqnarray}
The condition on $M_{l1}$ in (\ref{eq:(2)}) comes from the fact that the
corresponding vertex operator has charge equal to the electronic charge [see
(\ref{charge})] if we also use our assumption (\ref{kya}) that
$Q^{i}=e\delta _{i1}$.

It is easy to check that whereas the couplet $(G,B)$ of (\ref{4.28}) can
satisfy
(\ref{eq:(1)}) to (\ref{eq:(3)}), the couplet $(G''',B''')$ of (\ref{4.31})
cannot satisfy these equations (equation (\ref{eq:(2)}) being the one that
is incompatible with this couplet).  Thus there are no vertex operators with
the
charge of an electron here.  If we insist on the existence of excitations with
electronic charge (which is necessary if we want to identify the edge
wave functions with the bulk wave functions as done in Sections 2 and 3),
then we are forced to consider alternative sequences of $(G+B)$'s.

With this in mind, let us consider the sequence
of transformations given in (\ref{4.29}) but with a different starting
point :
\be
(G+B)^{-1} = \left( \begin{array}{cc}
m & 0 \\
2 & 2p \end{array} \right) , \label{onnu}
\ee
Then we have the following:
\be
(G+B)= \frac{1}{2pm} \left( \begin{array}{cc}
2p & 0 \\
-2 & m \end{array} \right)  ,            \label{rendu}
\ee
\be
(G'+B') = \left( \begin{array}{cc}
m & 0 \\
2 & 2p \end{array} \right)   ,             \label{moonu}
\ee
\be
(G''+B'') = \left( \begin{array}{cc}
m & 1 \\
1 & 2p \end{array} \right)    ,            \label{naalu}
\ee
\be
(G''' + B''') = \frac{1}{2pm-1}
\left( \begin{array}{cc}
2p & -1 \\
-1 & m
\end{array} \right)      .  \label{anju}
\ee
Thus we have a sequence similar to (\ref{haltra2}):
\be
\frac{1}{m} \stackrel{D}{\rightarrow} m \stackrel{E}{\rightarrow}
m \stackrel{D}{\rightarrow} \frac{1}{m - \frac{1}{2p}} \; \; .\label{aaru}
\ee
It can be checked now that (\ref{eq:(1)}) -(\ref{eq:(3)}) can be
satisfied for the starting matrix (\ref{rendu}) and the final matrix
(\ref{anju}).  For (\ref{anju}), it is trivial to check these relations because
it is a symmetric matrix (so that $G=G+B$).  We present a solution for the case
 where $(G+B)$ is given by     (\ref{rendu}).
One finds :
\begin{eqnarray}
N^{2} &=& 2p(N ^{1} +m) ,\nonumber\\
 M_{l1} &=& -1 ,\nonumber\\
 M_{l2} &=& \frac{(2pm-1)N^{1}}{2pm}+m ,\nonumber\\
 M_{1} &=& N^{1} /m ,\nonumber\\
 M_{2} &=& N^{1} +m - \frac{N^{1}}{pm}. \label{soln}
\end{eqnarray}
  Everything is
expressed in terms of $N^{1}$.  We see clearly that $pm$, and hence also $m$,
should be factors of $N^{1}$ if $M_{1}$ and $M_{2}$ are to be integers.
Furthermore,
one can check that $N.M$ is odd, as required, if
$\frac{(N^{1})^{2}}{m}=N^{1}(\frac{N^{1}}{m})$ is odd.  As $m$ divides $N^{1}$,
we thus have that $N^{1}$ and $m$ must be odd.  As $p$ divides the odd $N^{1}$,
$p$ too must be odd.  Thus both $m$ and $p$ have to be odd.  It is easy to
satisfy all these conditions.  Once they are all satisfied, $:e^{iM_{li}\phi
^{i}_{l}}$: is the electron operator.

Finally, in the case of $d$ scalars, by analogy with (\ref{onnu}), one can
start with
\be
(G+B)^{-1}=\left( \begin{array}{ccccc}
m & 0 & 0 & 0 & .. \\
2 & 2p_{1} & 0 & 0 & .. \\
0 & 2 & 2p_{2} & 0 & .. \\
0 & 0 & 2 & 2p_{3} & ..
\end{array} \right)         \label{4.1i}
\ee
instead of (\ref{4.1}) and get
\be
\mbox{The transformed }(G+B)^{-1}=\left( \begin{array}{ccccc}
m & 1 & 0 & 0 & .. \\
1 & 2p_{1} & 1 & 0 & .. \\
0 & 1 & 2p_{2} & 1 & .. \\
0 & 0 & 1 & 2p_{3} & 1..
\end{array} \right)          \label{4.1ti}
\ee
after the sequence of transformations $\tilde{E}^{(d-1)}*\tilde{E}^{(d-2)}*
\ldots \tilde{E}^{(1)}$. The transformed filling fraction will then be
\be
\label{zeenu}
\nu = \frac{1}{m-\frac{1}{2p_{1} - \frac{1}{2p_{2} ...}}}.
\ee

We also briefly note here that certain of Jain's fractions \cite{jain} can be
obtained with two scalar fields so that $d=2$.  They are given by
formulae very similar to what we have as the final filling fractions in
(\ref{haltra2}) or (\ref{aaru}) except that $m$ and $2p$ are interchanged.
Thus,
to obtain these fractions, there remains one more ``duality'' transformation
to be implemented after applying $\tilde{E}$ ($=D*E*D$) on the
couplet $(G,B)$
defined from (\ref{4.28}) or (\ref{onnu}).  The corresponding $O(2,2;{\bf Z})$
matrix of this last ``duality'' transformation is
\be
S=\left( \begin{array}{cccc} 0&1&0&0\\ 1&0&0&0\\ 0&0&0&1\\ 0&0&1&0 \end{array}
\right) \label{Simil}
\ee
Jain also has an additional hierarchy of fractions to arrive at which two other
operations are also needed.  They are
\begin{eqnarray}
&&\nu \rightarrow 1-\nu , \nonumber\\
&&\nu \rightarrow 1+\nu .\label{jai}
\end{eqnarray}
We, however, do not obtain these transformations of the filling fractions using
the duality transformations.

Nevertheless, we find it very intriguing
that most of the interesting filling fractions can be obtained using duality
transformations.  This implies in particular that the {\em spectra of edge
excitations at these different fractions are all identical} !
We do not yet have a good physical
argument explaining why this happens.  Clearly it is worth studying the
effects of all the $O(d,d;{\bf Z})$ generators in a more systematic way.

\sxn{Conclusions}
\nopagebreak
\par
	Let us summarize what has been done in this paper.  We have described
the edge states of a QHE sample by a multi-component massless scalar
field. The scalar fields are compactified on torii.  The zero modes thus live
on a periodic lattice with some ``metric'' $G$.  There is also an antisymmetric
matrix $B$
that describes a mixing of momenta and winding modes.  This theory is then
coupled to electromagnetism.  Since the edge currents of the Hall
system
are chiral, we impose
chirality on the currents of the scalar fields.  This makes the theory
anomalous and one has to add an extra term to the action to describe this
anomaly.
The coefficient of this term is proportional to the Hall conductivity
$\sigma _{H}$.  Actually, for reasons of gauge invariance, the full action also
has a Chern-Simons term with a uniquely determined overall coefficient
proportional to $\sigma _{H}$.  Now, when
we further require that there exist vertex operators describing electrons, we
find that $\sigma _{H}$ is forced to be a rational multiple of
$\frac{e^{2}}{2\pi}$.
We thus derive the quantization of $\sigma _{H}$ in a simple way.  By
analytically continuing the wave functions of the minimum energy
states at the boundary, we also precisely
get the Laughlin wave functions describing the ground states of the
multi-particle quantum theory in the interior of the disc.

We emphasize that while the calculations that fix the Hall conductivity are
being performed for the 1+1 dimensional field theory at the edge of
the disc, the considerations of Section 2 show that
there is a definite connection between the interior
physics that can be described by a CS theory and the conformal field theory
on the boundary \cite{witt,bal,wil}.  In fact it is this connection which
guarantees the chirality and gauge invariance of the edge current.
Indeed, a person `living' at the edge of the disk,
and unaware of anything in the interior, could have deduced many of the basic
features of the quantum Hall effect just
by requiring gauge invariance and chirality of the physical currents.

	Finally, and perhaps most interestingly, we find that $\sigma _{H}$ can
be transformed in well-defined ways by applying
generalized duality transformations, which are the familiar $O(d,d;{\bf Z})$
transformations of compactified string theories \cite{nar,kik,bala,dua}.  In
particular we can relate
the integer QHE's to the fractional QHE's including those that occur in most
of
the hierarchical schemes.  An important prediction we find is
that the {\em spectra of edge excitations at the fractions related by these
transformations are all identical.}  We find it rather remarkable that such
transformations should exist. A physical interpretation of their meaning would
be of interest for the study of the dynamics of the strongly correlated
electron
system in the interior of the disc.

\noindent
{\bf ACKNOWLEDGEMENTS}

We would also like to thank T.R.Govindarajan, A.Momen, John Varghese and
especially D.Karabali for many discussions.  A.P.B. also thanks C.Callan for
bringing reference \cite{nac} to his attention.

The work of APB and LC was supported by a grant from the Department of Energy
under contract number DE-FG02-85ER40231.

\end{document}